\documentclass[twocolumn,usenatbib]{aastex62}

\usepackage{latexsym}
\usepackage{longtable}
\usepackage{epsf}

\usepackage{graphicx}	% Including figure files
\usepackage[caption=false]{subfig}
\usepackage{amsmath}	% Advanced maths commands
\usepackage{amssymb}	% Extra maths symbols
\usepackage[shortlabels]{enumitem}
\usepackage{xcolor}

\definecolor{darkgreen}{RGB}{0,128,0}

\graphicspath{{./}{figures/}}

\shorttitle{Kilonova global asphericity}
\shortauthors{Darbha et al.}
\begin{document}

\title{Inclination Dependence of Kilonova Light Curves from Globally Aspherical Geometries}

\correspondingauthor{Siva Darbha}
\email{siva.darbha@berkeley.edu}

\author{Siva Darbha}
\affiliation{Department of Physics, University of California, Berkeley, Berkeley, CA 94720, USA}

\author{Daniel Kasen}
\affiliation{Department of Physics, University of California, Berkeley, Berkeley, CA 94720, USA}
\affiliation{Department of Astronomy and Theoretical Astrophysics Center, University of California, Berkeley, Berkeley, CA 94720, USA}
\affiliation{Nuclear Science Division, Lawrence Berkeley National Laboratory, Berkeley, CA 94720, USA}

%% Note that the \and command from previous versions of AASTeX is now
%% depreciated in this version as it is no longer necessary. AASTeX 
%% automatically takes care of all commas and "and"s between authors names.

%% AASTeX 6.2 has the new \collaboration and \nocollaboration commands to
%% provide the collaboration status of a group of authors. These commands 
%% can be used either before or after the list of corresponding authors. The
%% argument for \collaboration is the collaboration identifier. Authors are
%% encouraged to surround collaboration identifiers with ()s. The 
%% \nocollaboration command takes no argument and exists to indicate that
%% the nearby authors are not part of surrounding collaborations.

%% Mark off the abstract in the ``abstract'' environment. 
\begin{abstract}

% (The AAS Journals have a 250 word limit for the abstract.)

The merger of two neutron stars (NSs) or a neutron star and a black hole (BH)  produces a radioactively-powered transient known as a kilonova, first observed accompanying the gravitational wave event GW170817. While kilonovae are frequently  modeled in spherical symmetry, the dynamical ejecta and disk outflows can be considerably asymmetric. We use Monte Carlo radiative transfer calculations to study the light curves of kilonovae with globally axisymmetric geometries (e.g. an ellipsoid and a torus). We find that the variation in luminosity in these models is most pronounced at early times, and decreases until the light curves become isotropic in the late optically thin phase. The light curve shape and peak time are not significantly modified by the global asymmetry. We show that the projected surface area along the line of sight captures the primary geometric effects, and use this fact to provide a simple analytic estimate of the direction-dependent light curves of the aspherical ejecta. For the kilonova accompanying GW170817, accounting for asymmetry with an oblate (prolate) ellipsoid of axial ratio $2$ ($1/2$) leads to a $\sim 40 \%$ decrease (increase) in the inferred ejecta mass compared to the spherical case. The pole-to-equator orientation effects are expected to be significantly larger (a factor of $\sim 5 - 10$) for the more extreme asymmetries expected for some NS-BH mergers.

\end{abstract}

%% Keywords should appear after the \end{abstract} command. 
%% See the online documentation for the full list of available subject
%% keywords and the rules for their use.
\keywords{radiative transfer --- stars: neutron}

%% From the front matter, we move on to the body of the paper.
%% Sections are demarcated by \section and \subsection, respectively.
%% Observe the use of the LaTeX \label
%% command after the \subsection to give a symbolic KEY to the
%% subsection for cross-referencing in a \ref command.
%% You can use LaTeX's \ref and \label commands to keep track of
%% cross-references to sections, equations, tables, and figures.
%% That way, if you change the order of any elements, LaTeX will
%% automatically renumber them.
%%
%% We recommend that authors also use the natbib \citep
%% and \citet commands to identify citations.  The citations are
%% tied to the reference list via symbolic KEYs. The KEY corresponds
%% to the KEY in the \bibitem in the reference list below. 

\section{Introduction}
\label{sec:intro}

Neutron-rich matter ejected from a NS-NS or NS-BH merger can assemble into heavy elements via rapid neutron capture (the r-process) \citep{lattimer74,symbalisty82,eichler89,freiburghaus99}. The radioactive decay of the heavy elements produces a thermal transient known as a kilonova \citep{li98,metzger10}. The kilonova can be used to infer the properties of the merger, particularly when combined with the gravitational wave (GW) signal; this was shown explicitly in the recent binary NS merger event GW170817/AT2017gfo (for the GW event, see \citealt{abbott17a,abbott17b}; for the kilonova emission, see e.g. \citealt{abbott17b,arcavi17,chornock17,coulter17,cowperthwaite17,drout17,evans17,kasen17,kasliwal17,kilpatrick17,lipunov17,mccully17,nicholl17,pian17,shappee17,smartt17,soaressantos17,tanaka17,tanvir17,villar17}).

NS-NS mergers may eject mass via three general mechanisms \citep{metzger17}: tidal tails that are stripped from the tidally disrupted stars during inspiral, a shocked outflow that is expelled from the collision interface, and a disk wind emitted by the accretion disk formed after the merger. The first two mechanisms are referred to as the dynamical ejecta of the merger. In addition to these three  processes, NS-NS mergers may produce outflows driven by more particular or speculative mechanisms \citep{fujibayashi18,radice18b,metzger18,nedora19}.

The composition of an ejecta is determined by the electron fraction $Y_e = n_p / (n_n + n_p)$, where $n_p$ and $n_n$ are the number densities of protons/electrons and neutrons, respectively. Neutron-rich ejecta ($Y_e \lesssim 0.25$) can synthesize r-process material past the second r-process peak at $A \sim 130$, including the lanthanide species known to have very high opacities \citep{kasen13}. The emission when lanthanides are present has a redder spectrum and a later and broader peak \citep{kasen13,barnes13}.

Simulations of NS-NS mergers indicate that their ejecta are likely aspherical. The tidal tails lie primarily in the equatorial plane and have low $Y_e$ ($\lesssim 0.25$), and the shocked material lies primarily in a conical polar region and has a higher $Y_e$ due to neutrino and weak interactions (for early work on the tidal tails with Newtonian codes, see \citealt{rosswog99,rosswog00}; for more recent work on the tidal tails and collision ejecta with relativistic codes, see e.g. \citealt{hotokezaka13,bauswein13,radice16,radice18a,dietrich17}). The wind from the post-merger disk is fairly spherical, though mildly prolate, and likely comprised of a broad distribution of $Y_e$ \citep{perego14,siegel17,siegel18,fujibayashi18,radice18a,fernandez19}. The properties of the NSs and the binary system will influence the geometry and mass of each of these components. Mergers with more asymmetric mass ratios produce more massive tidal ejecta, and mergers with lower-radius NSs produce more massive shocked ejecta. More massive disk winds are produced by mergers with more asymmetric mass ratios or longer-lived NS remnants. The ejecta from NS-BH mergers is similar, but with two main differences: there is a single large tidal tail from the disruption of the NS only, and there is likely no collision interface ejecta since the BH does not have a material surface (e.g. \citealt{foucart13,foucart19,kyutoku13,kyutoku15,kawaguchi15,kiuchi15}).

The initial radiative transfer models of kilonovae assumed spherical symmetry \citep{metzger10,kasen13,barnes13}. These were used to fit the light curves and spectra of the event AT2017gfo and infer its basic properties (e.g. \citealt{chornock17,cowperthwaite17,kasen17,kilpatrick17,nicholl17,tanaka17,villar17}). Following this, several studies have examined aspherical kilonovae ejecta, using both radiative transfer simulations and semi-analytic models. \citet{roberts11} carried out 3D radiation transport simulations and found a factor of $\sim 3$ variation in the brightness with viewing angle. Subsequent radiation transport simulations that aimed to replicate the realistic, heterogeneous, multi-component ejecta structure have found similar results \citep{kasen15,kasen17,wollaeger18,kawaguchi18,kawaguchi19,bulla19}. 
% Recently, \citet{korobkin20} systematically studied one- and two-component axisymmetric models using an elegant set geometries based on Cassini ovals, and also found comparable results. 
\citet{grossman14} (and subsequently \citealt{perego14,martin15}) used a semi-analytic ``diffusive model'' to post-process the ejecta from merger simulations, and found a factor of $\sim 2$ variation in the brightness with viewing angle. \citet{barbieri19} have recently developed an autonomous semi-analytic model for multi-messenger parameter estimation, and derived a projection factor that yields a factor of $\sim 2 - 3$ variation.

A robust understanding of the dependence of kilonova light curves on geometry and inclination is needed to more accurately estimate the properties of kilonovae ejecta (e.g. mass, kinetic energy, opacities). In addition, a constraint on the inclination of the NS-NS merger from the kilonova would partly break the distance-inclination degeneracy in the GW data, and reduce the number of detections needed for an accurate Hubble constant measurement using joint kilonova/GW observations \citep{abbott17c,feeney19}. We note, though, that the inclination can also be measured by other means \citep{mooley18,hotokezaka19}, and the Hubble constant can be measured with the GW or kilonova data alone \citep{fishbach19,coughlin19}.

In this paper, we study the time-domain signatures from aspherical kilonovae produced by NS-NS and BH-NS mergers. In particular, we consider simple geometric models that describe the global behavior of kilonova ejecta to build intuition into how viewing angle effects depend on geometry. We focus on an ellipsoid and a ring torus, and extend our analysis to a conical section embedded in a sphere. We characterize the systematic uncertainties involved in using a spherical or aspherical model, and quantify the uncertainties as a function of sphericity. We find that the scale of the light curve variation with viewing angle is primarily determined by the projected surface area, and the light curves converge with time as they become more isotropic.

In Section \ref{sec:methods}, we outline the parameters and geometries of our ejecta models. In Section \ref{sec:results}, we present our results and provide a simple, semi-analytic prescription to estimate the direction-dependent light curve of the aspherical ejecta from the light curve of the equivalent spherical ejecta. In Section \ref{sec:discussion}, we discuss the range and limitations of our results, and apply them to a general conical geometry and to AT2017gfo. In Appendix \ref{sec:projected_area}, we provide equations for the parallel projected areas of our geometries. In Appendix \ref{sec:additional_features}, we present the fitting parameters, discuss the limitations of the semi-analytic prescription of Section \ref{sec:results}, and provide more involved parameterizations.

\section{Methods}
\label{sec:methods}

\subsection{General Properties}
\label{subsec:methods_general}

We study the emission from a homologously expanding ejecta using the time-dependent Monte Carlo radiative transfer code \textsc{sedona} \citep{kasen06}. We study an ejecta with mass $M$ and kinetic energy $E_k = Mv_\mathrm{ch}^2/2$, where $v_\mathrm{ch} = \beta_\mathrm{ch} c$ is the characteristic velocity and $c$ is the speed of light. We use a one-component, constant grey opacity $\kappa$ to parameterize the strength of the interaction between the thermal photons and matter.

The light curve evolution is determined by two timescales. The first is the effective diffusion timescale $t_d = (M\kappa / v_\mathrm{ch} c)^{1/2}$, which characterizes the time at which photons escape the ejecta faster than the ejecta can expand. We take this as our definition of $t_d$, but note that other definitions exist that contain additional numerical factors \citep{arnett82}. We use this to define the dimensionless time $\tau = t/t_d$, where $t$ is the physical time. The second is the thermalization timescale $t_e$, which characterizes the time at which the absorption of radioactive decay products begins to become inefficient. This can be written in terms of the dimensionless parameter $\tau_e = t_e/t_d$. For neutron-rich ejecta, the radioactive power from the r-process alone can be approximated by a power law \citep{metzger10,lippuner15,hotokezaka17}, and thus does not itself have an intrinsic timescale. We note that ejecta with $Y_e \gtrsim 0.35$ can exhibit a different time dependence \citep{wanajo14}.

The light curves are powered by the radioactive decay of neutron-rich elements synthesized by the r-process \citep{metzger10}. The fraction of the radioactive power that will thermalize and contribute to the EM emission is determined by its distribution in the different decay channels (beta, alpha, and fission) and their thermalization efficiencies \citep{metzger10,barnes16}, though the total contribution can be reduced to a simple approximate prescription \citep{kasen19}. We adopt the latter, yielding the specific heating rate
\begin{equation}
q(\tau;\tau_e) = q_0 t_{d,\mathrm{day}}^{\sigma_1} \tau^{\sigma_1} \left( 1 + \frac{\tau}{\tau_e} \right)^{\sigma_2}
\end{equation}
where $\tau_e$ is the thermalization timescale, $t_{d,\mathrm{day}}$ is the diffusion time in days, and we take $q_0 = 10^{10}$ ergs s$^{-1}$ g$^{-1}$, $\sigma_1 = -1.3$, and $\sigma_2 = -1.2$. The term $\tau^{\sigma_1}$ gives the time-dependence of the radioactive power from an ensemble of r-process nuclei \citep{metzger10,lippuner15,hotokezaka17}, and the term in parentheses is the thermalization efficiency \citep{kasen19}. We take $\tau_e = 10$ for our canonical example, but also examine the effects of different values. The total heating rate is then $Q = Mq$.

Given the above specific heating rate, we define the dimensionless luminosity $\Lambda = L/L_n$, where $L$ is the physical luminosity and $L_n = M q_0 t_{d,\mathrm{day}}^{\sigma_1}$ is the scale factor.  If all parts of the outflow are in the diffusive regime, then ejecta profiles with the same $\tau_e$ would be degenerate and exhibit the same dimensionless light curves $\Lambda(\tau)$, i.e. the light curve shape and anisotropy will not depend on the individual values of $M$, $\kappa$, $\beta_\mathrm{ch}$, and $q_0$, but only on the combinations $t_d = (M\kappa / \beta_\mathrm{ch} c^2)^{1/2}$ and $L_n = M q_0 t_{d,\mathrm{day}}^{\sigma_1}$ (as well as the variables $\tau_e$, $\sigma_1$, etc.). In practice, this scaling breaks down for outflows with a sufficiently high $\beta_\mathrm{ch}$ or low $M$, for which the outer ejecta layers are of low density and become optically thin. 
%Consequently, outflows with the same $\tau_e$ and $\beta_\mathrm{ch}$ are degenerate and exhibit the same dimensionless light curves, not those with the same $\tau_e$ alone. 
However, for a given $\tau_e$, we find that outflows with $\beta_\mathrm{ch} \lesssim 0.1$ generally remain in the diffusive regime.

With the above framework and qualifications, we run our simulations with fiducial ejecta parameters $M = 10^{-2} M_\odot$, $\beta_\mathrm{ch} = 0.1$, and $\kappa = 1$ cm$^2$ g$^{-1}$. We start our simulations at the initial time $\tau_0 = 0.01$ and nondimensionalize our results.

\subsection{Geometries}
\label{subsec:methods_geometries}

\begin{figure*}
\gridline{
\fig{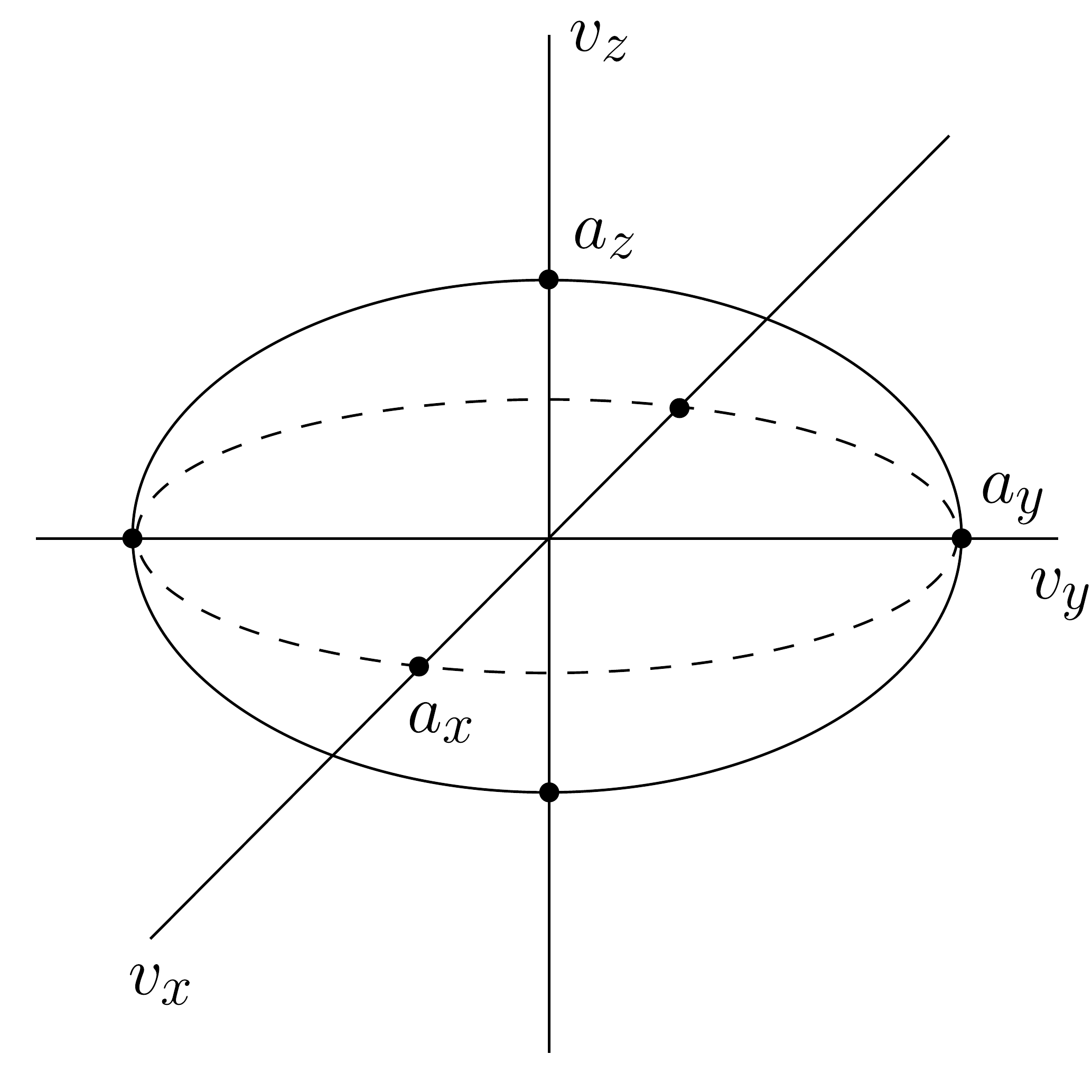}{0.32\textwidth}{(a) Ellipsoid}
\fig{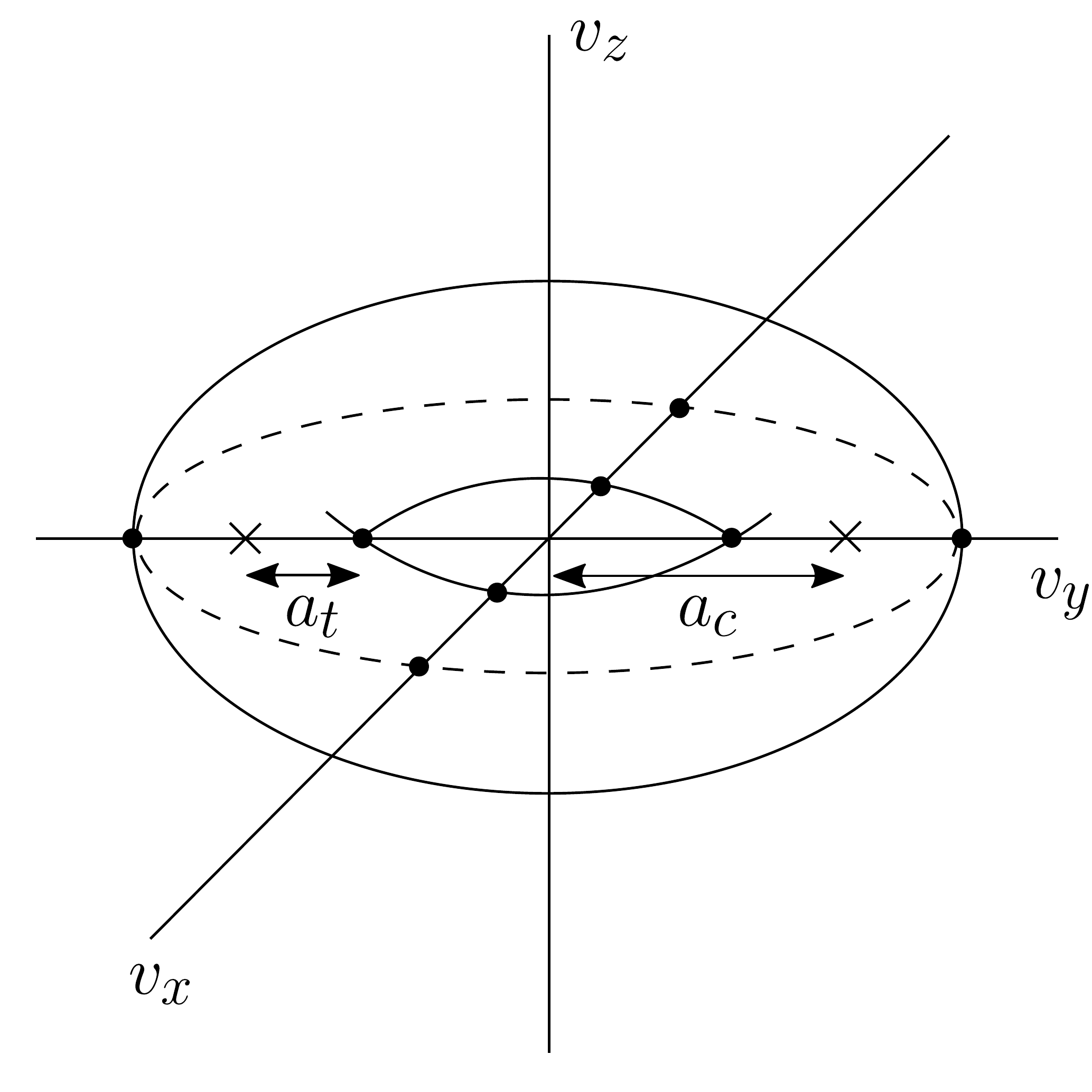}{0.32\textwidth}{(b) Torus}
\fig{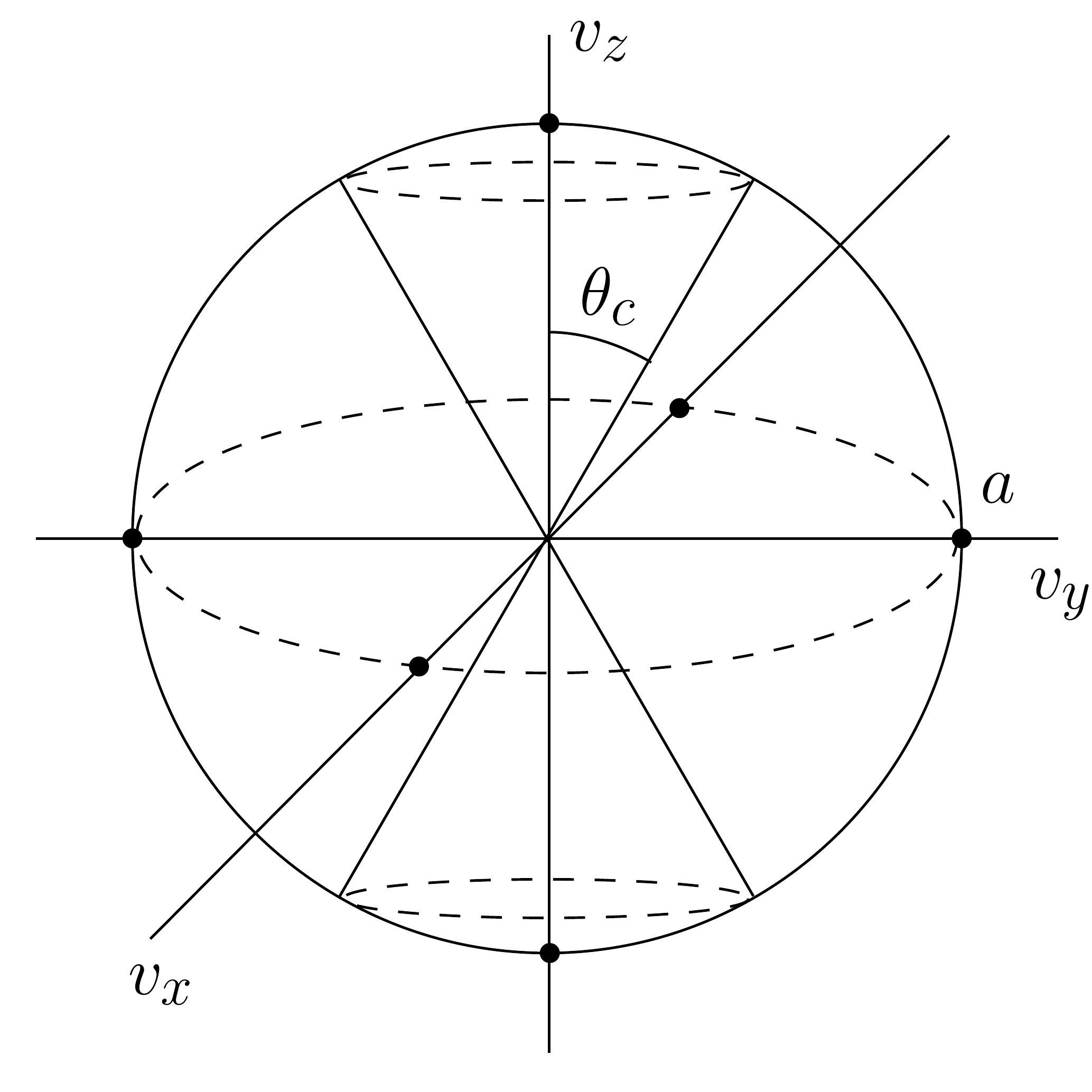}{0.32\textwidth}{(c) Conical Section}
}
\caption{The three ejecta geometries studied in this paper. The black dots show the points where the axes intersect the surfaces. a) An ellipsoid. The semi-major axes are $(a_x,a_y,a_z)$, and we study the axisymmetric case $a_x = a_y$ (spheroid) with axial ratios $R = a_x/a_z$ (oblate here, $R > 1$). (b) A torus. The spine radius is $a_c$ and the tube radius is $a_t$, and we study radius ratios $K = a_c/a_t > 1$ (ring torus). The crosses mark the points where the torus spine (i.e. the center of the tube) intersects the $v_y$-axis. (c) A conical section embedded in a sphere. The sphere has radius $a$ and the conical section has half-opening angle $\theta_c$.
}
\label{fig:geometries}
\end{figure*}

We study several idealized axisymmetric geometries that can represent the ejecta from a kilonova as described in Section \ref{sec:intro}. We summarize these geometries in Figure \ref{fig:geometries}. We primarily study an ellipsoid and a ring torus, and extend our analysis to a conical section embedded in a sphere.

These geometries are robust and versatile, and can replicate the shape of the different ejecta components described in Section \ref{sec:intro}. An ellipsoid is one of the simplest geometries that has dipolar asymmetry, and thus serves as a generic model for deviations from sphericity. A modestly prolate ellipsoid with an axial ratio $R$ (defined in Section \ref{subsubsec:geometries_ellipsoid}) near unity can serve as a model for the disk wind, which magnetohydrodynamics (MHD) simulations have shown can have a global distortion \citep{fernandez19}, with the amount of mass ejection varying by $\sim 2$ from pole to equator. An oblate ellipsoid with $R \gtrsim 2$ can serve as a model for the tidal tail, and a prolate ellipsoid with $R \lesssim 1/2$ can serve as a model for the collision-interface ejecta. A torus with a radius ratio $K$ (defined in Section \ref{subsubsec:geometries_torus}) near unity can serve as a model for the wind in the aftermath of jet puncture, and one with $K \gtrsim 2$ can also serve as a model for the tidal tail. A conical section embedded in a sphere can serve as a model for the low-opacity polar ejecta from the collision enshrouded by both the high-opacity equatorial ejecta from the tidal tail and the post-merger wind.

Since the geometries we study are axisymmetric, we parameterize the viewing angle with $\mu = \cos\theta$, where $\theta$ is the polar angle measured from the $z$-axis. We record the emission in $20$ uniformly space bins in $\mu \in [-1,1]$. In Appendix \ref{sec:projected_area}, we calculate the parallel projected areas as a function of $\mu$ for each of these geometries, which are useful for understanding the light curves.

\subsubsection{Ellipsoid}
\label{subsubsec:geometries_ellipsoid}

% We study an ellipsoid with velocity space semi-major axes of $(a_x,a_y,a_z)$, where $a_x = a_y$ (axisymmetric, spheroid), and we define the axial ratio $R = a_x/a_z$. We examine several axial ratios in the range $R \in [1,6]$. We parameterize the velocity profile of the ejecta as
% \begin{align}
% v_x &= a_x s \sin \theta \cos \phi \\
% v_y &= a_x s \sin \theta \sin \phi \\
% v_z &= a_z s \cos \theta
% \end{align}
% where the coordinates $(s,\theta,\phi)$ are dimensionless and lie in the range $0 \leq s \leq 1$, $0 \leq \theta \leq \pi$, $0 \leq \phi \leq 2\pi$. A surface of constant $s$ is the surface of an ellipsoid with its center at the origin and semi-major axes $(sa_x, sa_x, sa_z)$; the outer surface of the ejecta is given by $s=1$. The variables $(\theta,\phi)$ are the usual polar and azimuthal angles.

We study an ellipsoid with velocity coordinates $(v_x,v_y,v_z)$, and velocity space semi-major axes of $(a_x,a_y,a_z)$ where $a_x = a_y$ (axisymmetric, spheroid). We define the axial ratio $R = a_x/a_z$, and examine several values in the range $R \in [1/4,6]$. We define the dimensionless velocity coordinate
\begin{equation}
s = \left[ \left(\frac{v_x}{a_x}\right)^2 + \left(\frac{v_y}{a_y}\right)^2 + \left(\frac{v_z}{a_z}\right)^2 \right]^{1/2}
\end{equation}
where $0 \leq s \leq 1$. A surface of constant $s$ is the surface of an ellipsoid with its center at the origin and semi-major axes $(sa_x, sa_x, sa_z)$; $s=1$ corresponds to the outer surface of the ejecta.

We adopt a density profile of the form $\rho = \rho(s,\tau)$, in which surfaces of constant density correspond to surfaces of constant $s$. We study two types: 1) a constant density profile with a sharp cutoff at the surface, and 2) a broken power-law density profile. The constant profile is given by $\rho_c(s,\tau) = \rho_{c0}(\tau_0) (\tau_0/\tau)^3$. The broken power-law profile has been successfully used to model kilonovae and more general transients \citep{chevalier89,barnes13,kasen17}, and is given by
\begin{equation}
\rho_b(s,\tau) = \rho_{b0}(\tau_0) \left(\frac{\tau_0}{\tau}\right)^3
\begin{cases}
\left(\frac{s}{s_b}\right)^{\delta_1} ,& s < s_b \\
\left(\frac{s}{s_b}\right)^{\delta_2} ,& s_b \leq s \leq 1.
\end{cases}
\label{eq:density_profile}
\end{equation}
Here, $s_b$ is the location of the break in the power law, and $\delta_1$ and $\delta_2$ are the exponents in the broken power-law, which satisfy $0 > \delta_1 > -3$ and $\delta_1 > \delta_2$ to have a declining profile that decreases more precipitously after $s_b$ and has finite $M$ and $E_k$. We take $s_b = 0.5$, $\delta_1 = -1$, and $\delta_2 = -10$. We examine these two types of profiles in order to identify the trends in our results that are robust to changes in the form or parameters of the density profile. We use the values of $M$, $\beta_\mathrm{ch}$, and $\tau_0$ given in Section \ref{subsec:methods_general} and choose a value for $R$, and this sets the parameters $a_x$,  $a_z$, and either $\rho_{c0}(\tau_0)$ or $\rho_{b0}(\tau_0)$.

\subsubsection{Torus}
\label{subsubsec:geometries_torus}

% We study a torus with velocity space spine radius $a_c$ and tube radius $a_t$, where $a_c \geq a_t$ (ring torus), and we define the radius ratio $K = a_c/a_t \geq 1$. We examine several radius ratios in the range $K \in [1,5]$. We parameterize the velocity profile of the ejecta as
% \begin{align}
% v_x &= (a_c + a_t s \cos p) \cos q \\
% v_y &= (a_c + a_t s \cos p) \sin q \\
% v_z &= a_t s \sin p
% \end{align}
% where the coordinates $(s,p,q)$ are dimensionless and lie in the range $0 \leq s \leq 1$, $0 \leq p \leq 2\pi$, $0 \leq q \leq 2\pi$. A surface of constant $s$ is the surface of a torus with spine radius $a_c$ and tube radius $sa_t$; the outer surface of the ejecta is given by $s=1$. The variables $p$ and $q$ are the angles around the tube and the spine, respectively.

We study a torus with velocity coordinates $(v_x,v_y,v_z)$, and velocity space spine radius $a_c$ and tube radius $a_t$ where $a_c \geq a_t$ (ring torus). We define the radius ratio $K = a_c/a_t \geq 1$, and examine several values in the range $K \in [1,5]$. We define the dimensionless velocity coordinate
\begin{equation}
s = \frac{\left[ \left( (v_x^2 + v_y^2)^{1/2} - a_c \right)^2 + v_z^2 \right]^{1/2}}{a_t}
\end{equation}
where $0 \leq s \leq 1$. A surface of constant $s$ is the surface of a torus with spine radius $a_c$ and tube radius $sa_t$; $s=1$ corresponds to the outer surface of the ejecta.

We adopt the same types of density profile as in Section \ref{subsubsec:geometries_ellipsoid}, though with two modifications: the variable $s$ has the present definition, and the exponents satisfy $0 > \delta_1 > -2$ and $\delta_1 > \delta_2$. We take the same values for $s_b$, $\delta_1$, and $\delta_2$. We use the values of $M$, $\beta_\mathrm{ch}$, and $\tau_0$ given in Section \ref{subsec:methods_general} and choose a value for $K$, and this sets the parameters $a_c$,  $a_t$, and either $\rho_{c0}(\tau_0)$ or $\rho_{b0}(\tau_0)$.

\section{Results}
\label{sec:results}

The features of kilonovae light curves are determined by the properties of the underlying ejecta, and robust models are needed to extract the ejecta parameters from the observed light curves. In this section, we quantify the geometric effects that exist in aspherical ejecta, and present a simple and intuitive method to roughly estimate the anisotropic light curves of aspherical kilonovae given the light curve of the equivalent spherical model (with the same $M$, $E_k$, and $\tau_e$).

We first study an ellipsoid as a generic model for dipolar deviations from sphericity. The bolometric light curves from this geometry serve as a surrogate for the frequency-integrated (e.g. blue or red) unobstructed light curves of the different ejecta components after reintroducing the appropriate scales. We focus on a broken power-law density profile, and highlight the differences in the constant density case. We then examine a torus, and describe the geometry-specific differences compared to the ellipsoid. Appendix \ref{sec:additional_features} presents the values of various fitting parameters and approximations used in our prescriptions for both geometries and density profiles. The numerical models are also available if accurate light curves are needed.

\subsection{Ellipsoid}
\label{subsec:results_ellipsoid}

\begin{figure*}
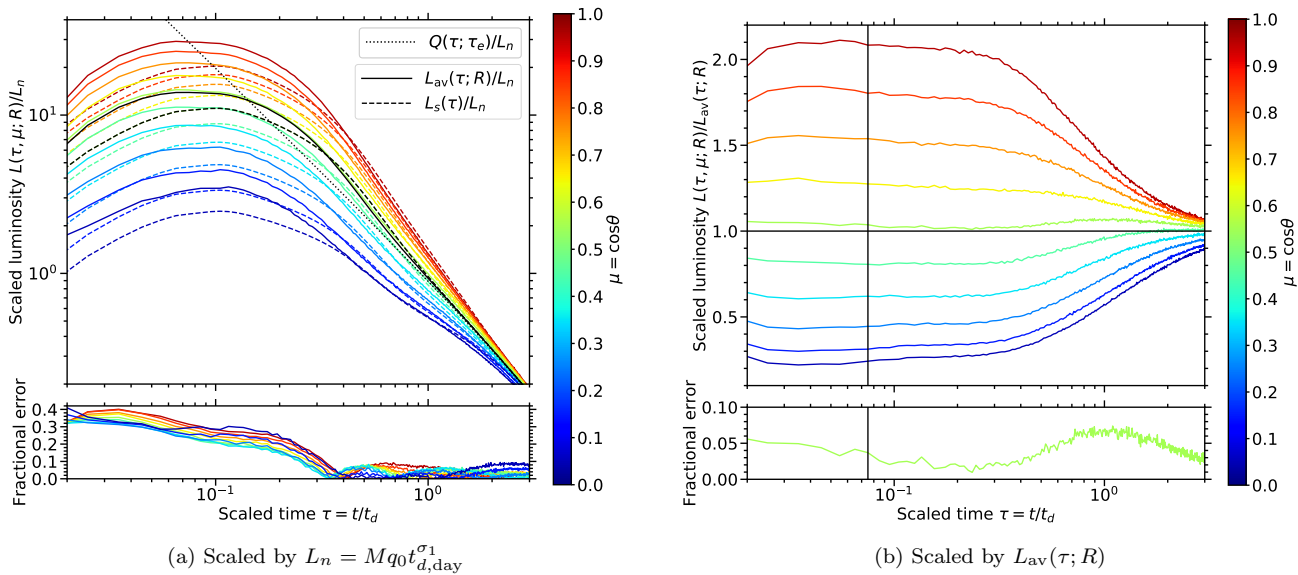

\gridline{
\fig{{L_of_tau_mu_vs_tau-with_curve_fits-g_1-R_4}.pdf}{0.46\textwidth}{(a) Scaled by $L_n = M q_0 t_{d,\mathrm{day}}^{\sigma_1}$}
\fig{{h_of_tau_mu_vs_tau-R_4}.pdf}{0.46\textwidth}{(b) Scaled by $L_\mathrm{av}(\tau;R)$}
}
\caption{
The isotropic-equivalent bolometric light curves $L(\tau,\mu;R)$ for an ellipsoid outflow with a broken power-law density profile and axial ratio $R = 4$ at different viewing angles $\mu = \cos\theta$. The time $t$ has been scaled by the diffusion time $t_d$ as $\tau = t / t_d$, and the thermalization time is $\tau_e = 10$. (a) The isotropic-equivalent luminosity has been scaled by the factor $L_n = M q_0 t_{d,\mathrm{day}}^{\sigma_1}$ (Section \ref{subsec:methods_general}). The solid black curve shows the angle-averaged luminosity, the dashed black curve shows the luminosity of the equivalent spherical ejecta ($R=1$ with the same $\tau_e$), and the dotted black curve shows the total heating rate. The dashed colored curves show the curve fits using the functions in Equations \ref{eq:L_of_tau_mu_R_analytic} and \ref{eq:fitting_function_k}. The bottom panel shows the error of the fits. (b) The isotropic-equivalent luminosity has been scaled by the angle-averaged luminosity $L_\mathrm{av}(\tau;R)$. The solid black horizontal curve shows the angle-averaged luminosity, and the solid black vertical curve shows its time to peak. The bottom panel shows the error of $L(\tau,\mu_\mathrm{ref};R)$ from $L_\mathrm{av}(\tau;R)$ for $\mu_\mathrm{ref} = 0.55$.
}
\label{fig:scaled_lightcurves-ellipsoid-bpl_density}
\end{figure*}

Figure \ref{fig:scaled_lightcurves-ellipsoid-bpl_density} shows the isotropic-equivalent light curves $L(\tau,\mu;R)$ for the ellipsoidal kilonova with a broken-power law density profile and an axial ratio $R = 4$. The viewing angle is defined by $\mu = \cos\theta$ (where $\theta$ is the polar angle measured from the $z$-axis) and each value of $\mu$ has equal probability of being observed. The light curves are invariant under the transformation $\mu \rightarrow -\mu$ due to reflection symmetry about $z=0$, so the figure shows only the viewing angles $\mu > 0$. 

The light curves of the ellipsoidal model are brightest along the pole ($\theta = 0$, or $\mu = 1$) and dimmest along the equator ($\theta = \pi/2$, or $\mu = 0$). The total variation in the luminosity at peak is about a factor of $\sim 9$. The time to peak $\tau_p(\mu;R) \sim 0.1$, though, remains largely unchanged with viewing angle. The viewing angle dependence is most pronounced at early times, but decreases over time until the light curves become mostly isotropic by $\tau \sim 5$ and converge to the total heating rate $Q(\tau;\tau_e)$. This is because at late times the ejecta becomes fully optically thin, so that all parts of the ejecta radiate equally in all directions (apart from small relativistic corrections).

Figure \ref{fig:scaled_lightcurves-ellipsoid-bpl_density} also shows the angle-averaged light curve $L_\mathrm{av}(\tau;R)$ of the ellipsoidal model, and the isotropic-equivalent light curves normalized to this. The light curve of an intermediate viewing angle $\mu_\mathrm{ref} \approx 0.55$ ($\theta_{\rm ref} \approx 57^\circ$) roughly equals the angle-averaged light curve, $L(\tau,\mu_\mathrm{ref};R) \simeq L_\mathrm{av}(\tau;R)$, with an error $\epsilon < 0.1$. The polar light curves are brighter than these and the equatorial ones are dimmer. The angle-averaged light curve is comparable to the light curve $L_s(\tau)$ of the equivalent spherical model ($R=1$ with the same $\tau_e$), $L_\mathrm{av}(\tau;R) \simeq L_s(\tau)$, with an error $\epsilon < 0.5$ (Figure \ref{fig:g_of_tau_vs_tau-ellipsoid}).

The basic viewing angle dependence of the light curves can be understood by considering the parallel projected surface area of the ejecta at different inclinations. If we consider a simple approximation in which an ellipsoidal photospheric shell emits blackbody radiation at a fixed temperature $T_{bb}$, then the luminosity would be proportional to the projected area, which is found to be (Appendix \ref{subsec:projected_area_ellipsoid})
\begin{equation}
A_\mathrm{proj}(\mu;R) = \pi R a_z^2 [ (R^2 - 1) \mu^2 + 1 ]^{1/2}
\label{eq:projected_area-ellipsoid}
\end{equation}
For an oblate ellipsoid, the projected area is a maximum along the pole and decreases monotonically to a minimum at the equator. The pole-to-equator projected area ratio is
\begin{equation}
\frac{A_\mathrm{proj}(\mu=1;R)}{A_\mathrm{proj}(\mu=0;R)} = R
\label{eq:pole_to_equator_ratio-ellipsoid}
\end{equation}
and this gives a rough scale for the luminosity variation of the light curve from pole to equator.

In reality, the kilonova ejecta is not described by a constant temperature ellipsoidal photosphere. Rather, the photosphere recedes with time and there will be emission from a surface of non-constant temperature plus emission from the optically thin volume outside the photosphere. The different viewing angles will have different effective diffusion times and effective photospheric parameters. Nevertheless, we show below that the projected area appears to capture the primary geometric effects on the light curve.

\begin{figure*}
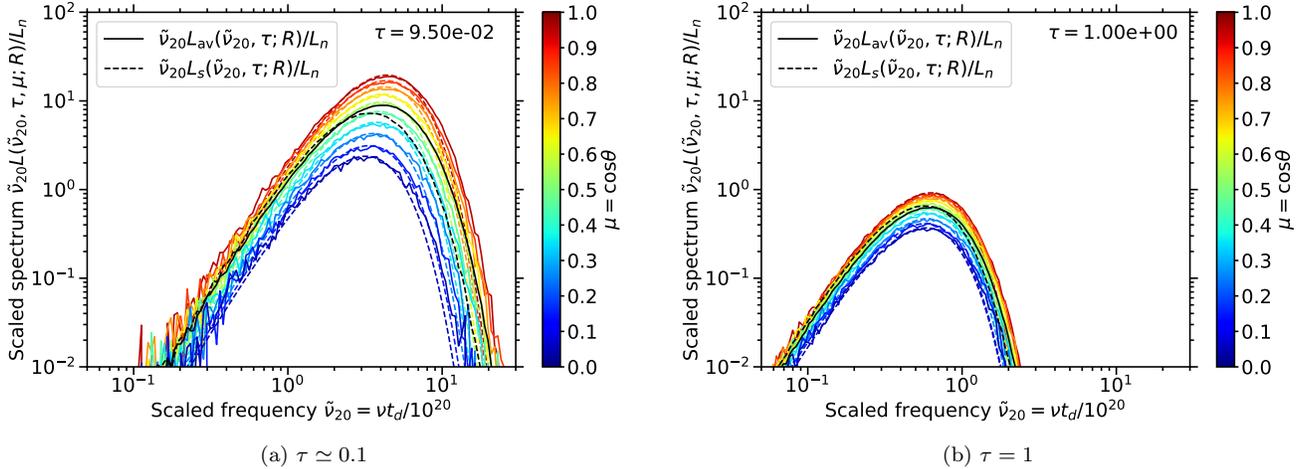

\gridline{
\fig{{nu_Lnu_vs_nu-legend_mu-t_9.50e-02-with_curve_fits-R_4}.pdf}{0.46\textwidth}{(a) $\tau \simeq 0.1$}
\fig{{nu_Lnu_vs_nu-legend_mu-t_1.00e+00-with_curve_fits-R_4}.pdf}{0.46\textwidth}{(b) $\tau = 1$}
}
\caption{
The spectra for an ellipsoid outflow with a broken power-law density profile and axial ratio $R = 4$ at different viewing angles and times. The time $t$ has been scaled by the diffusion time $t_d$ as $\tau = t / t_d$, the frequency $\nu$ has been scaled as $\tilde{\nu}_{20} = \nu t_d / 10^{20}$, and the luminosity $L(\tilde{\nu}_{20},\tau,\mu;R)$ has been scaled by $\tilde{\nu}_{20}$ and $L_n = M q_0 t_{d,\mathrm{day}}^{\sigma_1}$ (Section \ref{subsec:methods_general}). The spectra are shown at times (a) $\tau \simeq 0.1$ and (b) $\tau = 1$. The solid black curves show the angle-averaged spectra and the dashed black curves show the spectra of the equivalent spherical ejecta. The dashed colored curves show the curve fits to an effective blackbody (Equation \ref{eq:blackbody_luminosity}).
}
\label{fig:nu_Lnu_vs_nu-with_curve_fits}
\end{figure*}

Figure \ref{fig:nu_Lnu_vs_nu-with_curve_fits} shows the spectra as a function of viewing angle at two different times for $R=4$. The spectra for different $\mu$ are well-described by an effective blackbody
\begin{equation}
L(\nu,\tau,\mu;R) = 4\pi R_\mathrm{surf}^2(\tau) B(\nu,T_\mathrm{eff}(\tau))
\label{eq:blackbody_luminosity}
\end{equation}
where $\nu$ is the photon frequency and $B$ is the blackbody spectrum
\begin{equation}
B(\nu, T_\mathrm{eff}) = \frac{2 h \nu^3 / c^2}{e^{h \nu / k T_\mathrm{eff} - 1}}
\label{eq:blackbody_flux}
\end{equation}
with effective temperature $T_\mathrm{eff}$ and effective areal radius $R_\mathrm{surf}$. The blackbody form arises since we used a constant grey opacity (Section \ref{subsec:methods_general}). 
% We can write the spectra as
% \begin{equation}
% \tilde{\nu}_{20} L(\tilde{\nu}_{20},\tau,\mu;R) = 4\pi \tilde{R}_\mathrm{surf}^2 \frac{2 h (10^{20} \tilde{\nu}_{20})^4 / t_d^2}{e^{\tilde{\nu}_{20} / \tilde{T}_{\mathrm{eff},20}} - 1}
% \label{eq:effective_blackbody}
% \end{equation}
% where we use the dimensionless variables $\tilde{\nu}_{20} = \nu t_d / 10^{20}$, $\tilde{T}_{\mathrm{eff},20} = k_B T_\mathrm{eff} t_d / (10^{20} h)$, and $\tilde{R}_\mathrm{surf} = R_\mathrm{surf} / (c t_d)$. 
The effective temperatures and radii increase with projected surface area: they are larger for $\mu$ towards the poles and smaller for $\mu$ towards the equator. The temperatures decrease and the radii increase with time until $\tau \sim 2$; the temperatures then converge to one constant value and the radii decrease exponentially with time while retaining their dependence on $\mu$. The viewing angle variation of the spectra thus decreases with time, as expected from the light curves.

\begin{figure}
\includegraphics[width=0.46\textwidth]{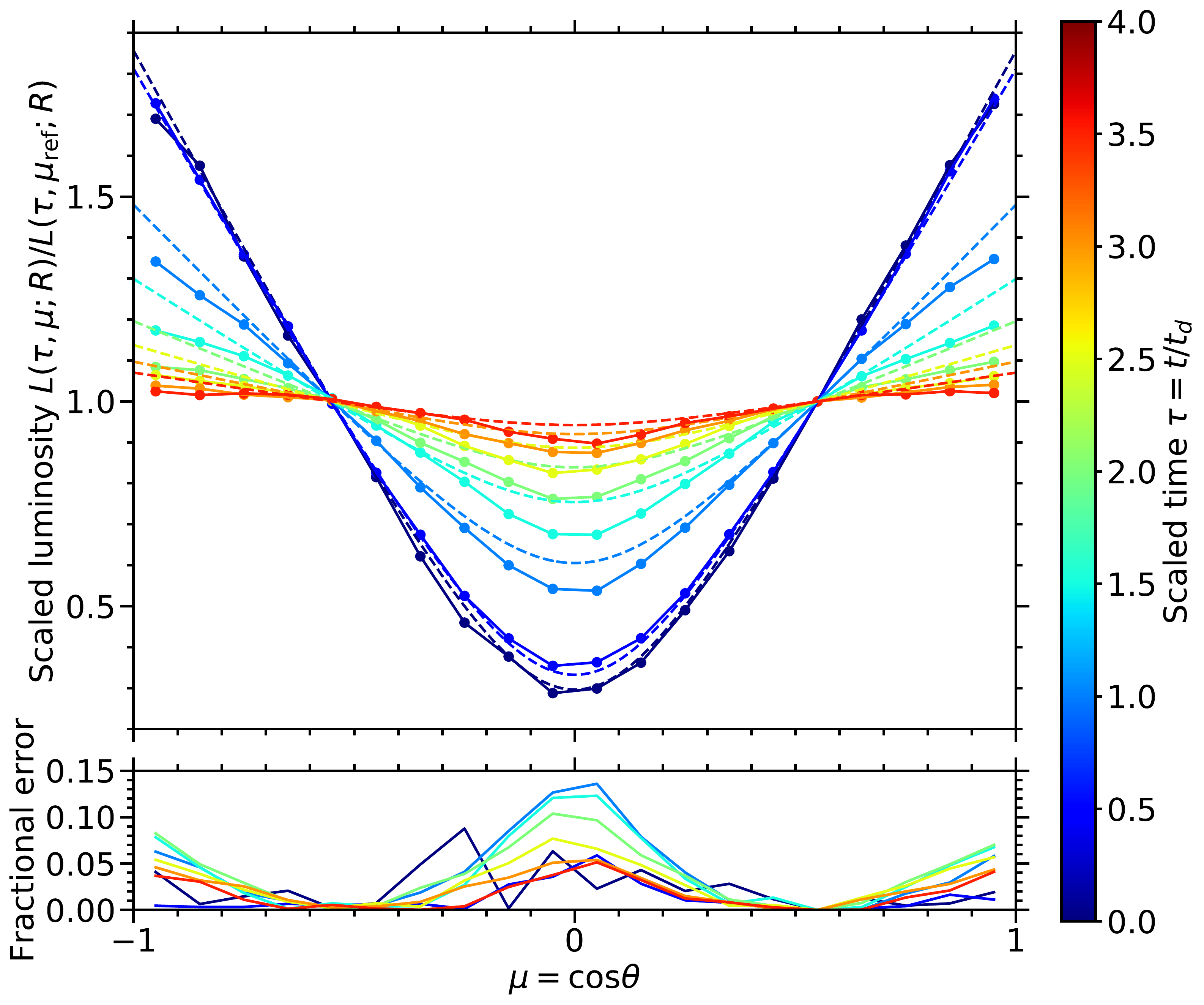}
\caption{The isotropic-equivalent bolometric luminosities for an ellipsoid outflow with a broken power-law density profile and axial ratio $R = 4$ as a function of viewing angle $\mu = \cos \theta$ and at different times. The time $t$ has been scaled by the diffusion time $t_d$ as $\tau = t / t_d$ (Section \ref{subsec:methods_general}), and the isotropic-equivalent luminosity $L(\tau,\mu;R)$ has been scaled by that in the reference direction $\mu_\mathrm{ref} = 0.55$, since in this direction the luminosity roughly equals the angle-averaged luminosity. The thermalization time is $\tau_e = 10$. The points and solid colored curves show the luminosities at different values of $\tau < 4$. The dashed colored curves show the fits to the function in Equation \ref{eq:L_of_tau_mu_R_analytic}. The bottom panel shows the error of the fits.}
\label{fig:psi_of_tau_mu_vs_mu-ellipsoid}
\end{figure}

Figure \ref{fig:psi_of_tau_mu_vs_mu-ellipsoid} shows the variation of the luminosity ratio $L(\tau,\mu;R) / L(\tau,\mu_\mathrm{ref};R)$ with viewing angle at different times for $R=4$. At the time $\tau \simeq 0.65$ (not shown in the figure), the angular dependence is remarkably well fit by the ratio of projected surface areas $A_\mathrm{proj}(\mu;R) / A_\mathrm{proj}(\mu_\mathrm{ref};R)$ (Equation \ref{eq:projected_area-ellipsoid}). At other times, the viewing angle dependence has the same shape (Figure \ref{fig:projected_area}), but with a different scale. This suggests that we can roughly describe the basic behavior of the light curve with a simple time-dependent function of the projected area (Equation \ref{eq:projected_area-ellipsoid}), such as
\begin{equation}
\frac{L(\tau,\mu; R)}{L(\tau,\mu_\mathrm{ref};R)} \simeq 1 + k(\tau;R) \left( \frac{A_\mathrm{proj}(\mu;R)}{A_\mathrm{proj}(\mu_\mathrm{ref};R)} - 1 \right)
\label{eq:L_of_tau_mu_R_analytic}
\end{equation}
where $k(\tau;R)$ is a dimensionless fitting parameter that describes the scale of the viewing angle variation as a function of time. The dashed lines in Figure \ref{fig:psi_of_tau_mu_vs_mu-ellipsoid} show the fits to $L(\tau,\mu;R) / L(\tau,\mu_\mathrm{ref};R)$ using this function.

The fitting parameter $k(\tau;R)$ rises with time to a peak around $\tau \sim 0.3$ and then decays to zero roughly exponentially (Figure \ref{fig:k_vs_tau-ellipsoid}). We can see its effect in Figure \ref{fig:scaled_lightcurves-ellipsoid-bpl_density}, which shows the ratio $L(\tau,\mu;R)/L_\mathrm{av}(\tau;R)$ as a function of time for $R=4$. The curves show that the orientation effects are largest at early times, when the ejecta is optically thick, and converge at late times, when the ejecta becomes optically thin and the emission becomes isotropic (apart from Doppler shift effects, which are small for the values of $\beta_\mathrm{ch}$ of interest). The effects fall off roughly exponentially with time.

A more detailed inspection suggests that we can approximate $k(\tau;R)$ with the analytic expression
\begin{equation}
k(\tau;R) \simeq k_0(R) \frac{2 + \tau/\tau_b(R)}{1 + e^{\tau/\tau_b(R)}}
\label{eq:fitting_function_k}
\end{equation}
with the fitting parameters $k_0(R)$ and $\tau_b(R)$, where $k_0(R)$ sets the value at $\tau = 0$ and $\tau_b(R)$ is an estimate of the time scale for the anisotropy to decay. At early times $\tau \ll \tau_b(R)$, the expression becomes $k(\tau;R) \sim k_0(R)$ with $\partial_\tau k = 0$. At late times $\tau \gg \tau_b(R)$, it becomes $k(\tau;R) \sim \tau e^{-\tau/\tau_b(R)}$, which is close to the observed exponential fall off. Using this expression for $R=4$, we find the values $k_0 = 1.4$ and $\tau_b = 0.59$, and the fit residual $|\delta| < 0.1$ (Figure \ref{fig:k_vs_tau-ellipsoid}).

We can thus obtain a rough projection factor from $L_s(\tau)$ to $L(\tau,\mu;R)$ by using Equation \ref{eq:L_of_tau_mu_R_analytic}, with the replacement $L(\tau,\mu_\mathrm{ref};R) \rightarrow L_s(\tau)$, and Equation \ref{eq:fitting_function_k}, with the values given for $k_0$ and $\tau_b$. Figure \ref{fig:scaled_lightcurves-ellipsoid-bpl_density} shows the semi-analytic light curves for $R = 4$ obtained by applying this projection factor, along with their errors. The projected light curves are not impeccable; they typically have lower peaks and converge more quickly to the heating rate at late times. However, they are accurate to within an error of $\epsilon \lesssim 0.4$ before peak and $\epsilon \lesssim 0.3$ after peak, and thus roughly capture the temporal evolution of the emission and the correct scale of the variation.

The projection factor obtained from this approach provides a workable estimate of the viewing angle dependence, though it has limitations and should not be used cavalierly. Firstly, the fitting parameter $k(\tau;R)$ rises to a peak and decays, and is bounded above since Equation \ref{eq:L_of_tau_mu_R_analytic} must be positive; in contrast, the analytic approximation in Equation \ref{eq:fitting_function_k} starts at a peak with zero slope and decreases monotonically with increasing $\tau$. 
% , and must be used with care to ensure that it does not surpass the upper bound on $k(\tau;R)$. 
Secondly, the angle-averaged light curves $L_\mathrm{av}(\tau;R)$ are not quite equal to the equivalent spherical light curves $L_s(\tau)$, and the divergence between the two is more pronounced for larger axial ratios. Appendix \ref{sec:additional_features} quantifies the scale of these effects, and provides more accurate but more involved parameterizations. 
% The numerical models are also available if accurate light curves are needed.

The features presented above for $R=4$ generalize over our range of $R \in [0.25,6]$. Ejecta with higher $R$ have a larger spread in brightness since the projected area changes more from pole to equator (Equation \ref{eq:pole_to_equator_ratio-ellipsoid}). The time to peak $\tau_p(\mu;R)$ is largely insensitive to $R$ in addition to $\mu$, and falls within $0.05 \leq \tau_p(\mu;R) \leq 0.12$. The various fits and approximations are more accurate for $R$ closer to unity and grow less accurate with increasing asphericity. The relation $L(\tau,\mu_\textrm{ref};R) \simeq L_\mathrm{av}(\tau;R)$ for the reference value $\mu_\mathrm{ref} = 0.55$ has errors $\epsilon \lesssim 0.1$. The relation $L_\mathrm{av}(\tau;R) \simeq L_s(\tau)$ has error $\epsilon < 0.1$ for $0.25 \leq R \leq 2$, rising to $\epsilon < 0.9$ for $R = 6$ (Figure \ref{fig:g_of_tau_vs_tau-ellipsoid}). For different $R$, the fitting parameter $k(\tau;R)$ has a similar time evolution and the analytic approximation for $k(\tau;R)$ remains robust (Figure \ref{fig:k_vs_tau-ellipsoid}). The semi-analytic light curves have errors $\epsilon < 0.2$ for $0.25 \leq R \leq 2$, rising to $\epsilon < 0.6$ for $R = 6$, which are comparable to those from uncertainties in the opacity \citep{barnes13} and heating rate \citep{metzger10,korobkin12,lippuner15}. More simply, we find that the constant values $k_0(R) \simeq 1.2$ and $\tau_b(R) \simeq 0.7$ produce semi-analytic light curves within these error bounds.

\begin{figure}
\includegraphics[width=0.46\textwidth]{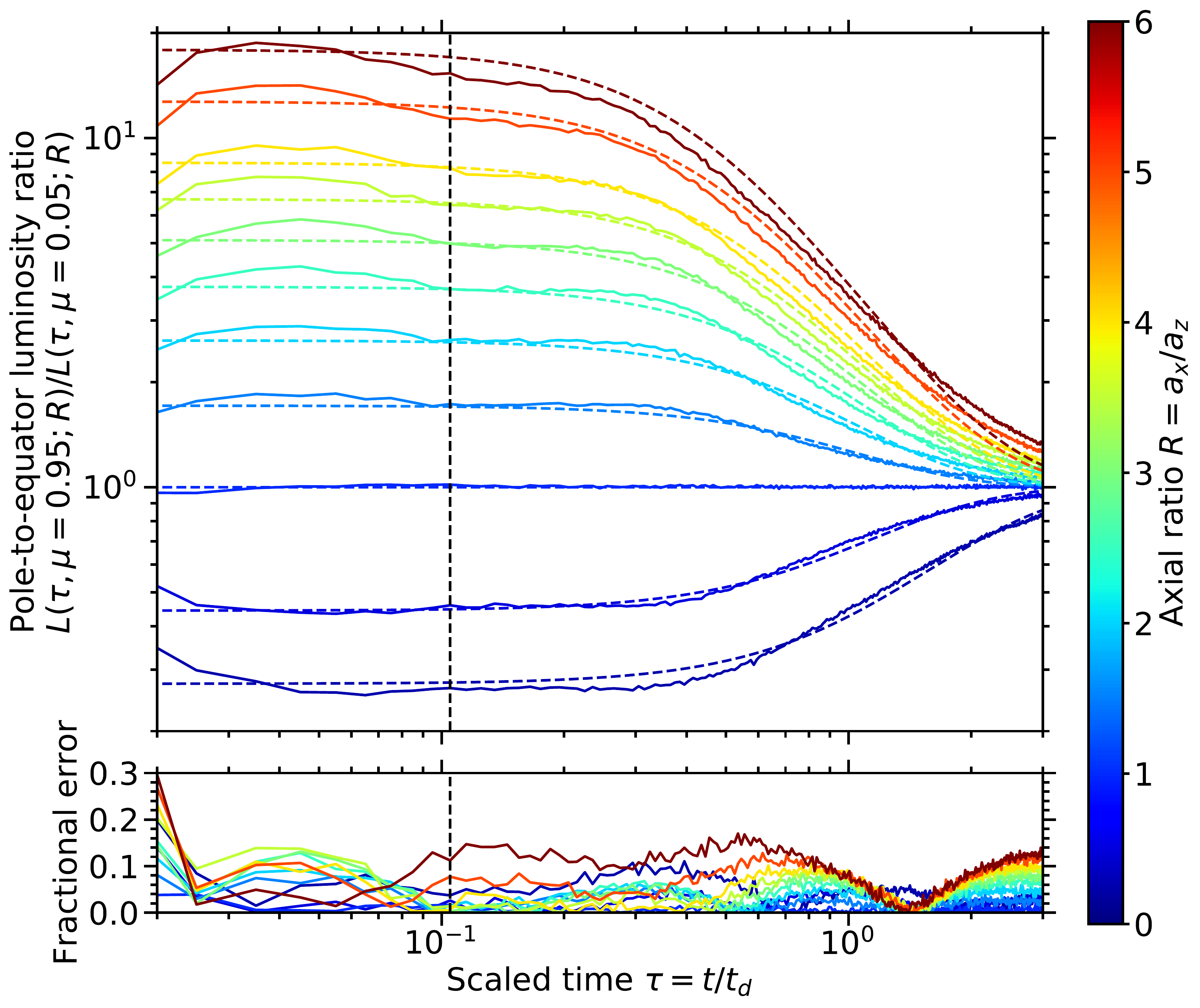}
\caption{The pole-to-equator light curve ratio for an ellipsoid outflow with a broken power-law density profile and different axial ratios $R$. The time $t$ has been scaled by the diffusion time $t_d$ as $\tau = t / t_d$ (Section \ref{subsec:methods_general}), and the isotropic-equivalent luminosity $L(\tau;\mu,R)$ roughly takes a polar value in the direction $\mu = 0.95$ and an equatorial value in the direction $\mu = 0.05$. The thermalization time is $\tau_e = 10$. The dashed black vertical curve shows the time to peak of the light curve of the equivalent spherical ejecta ($R=1$ with the same $\tau_e$). The dashed colored curves show the fits using the functions in Equations \ref{eq:L_of_tau_mu_R_analytic} and \ref{eq:fitting_function_k}. The bottom panel shows the error of the fits.}
\label{fig:Lpole_over_Lequator_vs_tau}
\end{figure}

We summarize the inclination dependence for different $R$ with the pole-to-equator luminosity ratio, shown in Figure \ref{fig:Lpole_over_Lequator_vs_tau}. The curves inherit the behavior described in Figures \ref{fig:scaled_lightcurves-ellipsoid-bpl_density} and \ref{fig:psi_of_tau_mu_vs_mu-ellipsoid} and the related text. In particular, at early times the luminosity ratio is larger than the corresponding projected area ratio ($\sim R$, Equation \ref{eq:pole_to_equator_ratio-ellipsoid}), and at late times it approaches unity as the emission becomes isotropic. The dashed colored curves show fits using the analytic approximations in Equations \ref{eq:L_of_tau_mu_R_analytic} and \ref{eq:fitting_function_k} with the fitting parameters $k_0(R)$ and $\tau_b(R)$ in Figure \ref{fig:k_vs_tau-ellipsoid}. The fits are accurate within errors $\epsilon < 0.3$ during the rise phase $\tau \lesssim \tau_p(\mu;R) \sim 0.1$, and within $\epsilon < 0.2$ after this.

\begin{figure*}
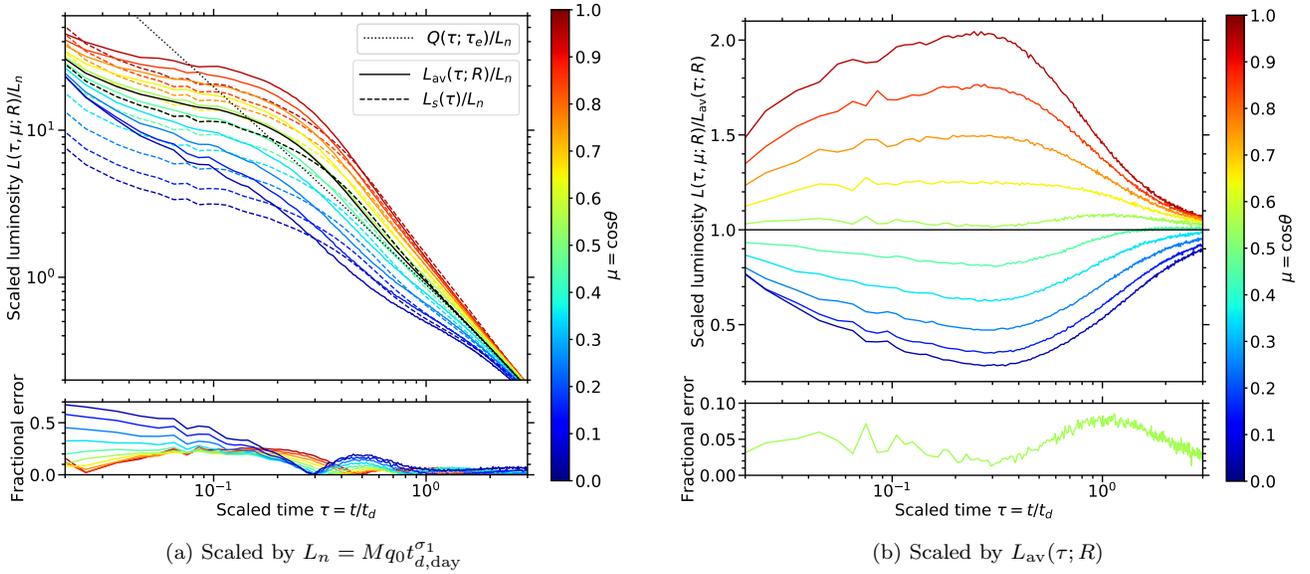

\gridline{
\fig{{L_of_tau_mu_vs_tau-with_curve_fits-g_1-R_4-constant_density}.pdf}{0.46\textwidth}{(a) Scaled by $L_n = M q_0 t_{d,\mathrm{day}}^{\sigma_1}$}
\fig{{h_of_tau_mu_vs_tau-R_4-constant_density}.pdf}{0.46\textwidth}{(b) Scaled by $L_\mathrm{av}(\tau;R)$}
}
\caption{
The isotropic-equivalent bolometric light curves $L(\tau,\mu;R)$ for an ellipsoid outflow with a constant density profile and axial ratio $R = 4$ at different viewing angles $\mu = \cos\theta$. The time $t$ has been scaled by the diffusion time $t_d$ as $\tau = t / t_d$, and the thermalization time is $\tau_e = 10$. (a) The isotropic-equivalent luminosity has been scaled by the factor $L_n = M q_0 t_{d,\mathrm{day}}^{\sigma_1}$ (Section \ref{subsec:methods_general}). The solid black curve shows the angle-averaged luminosity, the dashed black curve shows the luminosity of the equivalent spherical ejecta ($R=1$ with the same $\tau_e$), and the dotted black curve shows the total heating rate. The dashed colored curves show the curve fits using the functions in Equations \ref{eq:L_of_tau_mu_R_analytic} and \ref{eq:fitting_function_k}. The bottom panel shows the error of the fits. (b) The isotropic-equivalent luminosity has been scaled by the angle-averaged luminosity $L_\mathrm{av}(\tau;R)$. The solid black horizontal curve shows the angle-averaged luminosity. The bottom panel shows the error of $L(\tau,\mu_\mathrm{ref};R)$ from $L_\mathrm{av}(\tau;R)$ for $\mu_\mathrm{ref} = 0.55$.
}
\label{fig:scaled_lightcurves-ellipsoid-constant_density}
\end{figure*}

The constant density case is similar to the broken power-law case, with some systematic differences. Figure \ref{fig:scaled_lightcurves-ellipsoid-constant_density} shows the isotropic-equivalent light curves for a constant density profile and an axial ratio $R = 4$, as an example. The light curves begin at a peak and decline monotonically since the constant density ejecta has more mass at larger distances. The pole-to-equator projected area ratio again gives the rough scale of the luminosity variation at intermediate times. The relation $L(\tau,\mu_\textrm{ref};R) \simeq L_\mathrm{av}(\tau;R)$ has errors $\epsilon \lesssim 0.1$ for the reference value $\mu_\mathrm{ref} = 0.55$. This suggests that $\mu_\mathrm{ref}$ is a function primarily of the geometry. The relation $L_\mathrm{av}(\tau;R) \simeq L_s(\tau)$ has error $\epsilon < 0.1$ for $0.25 \leq R \leq 2$, rising to $\epsilon < 0.4$ for $R = 6$ (Figure \ref{fig:g_of_tau_vs_tau-ellipsoid}). The fitting parameters $k(\tau;R)$ begin at lower values at early time but exhibit similar peak values and decline rates as the broken power-law case, and the analytic approximation for $k(\tau;R)$ has similar parameters (Figure \ref{fig:k_vs_tau-ellipsoid}). The projection factor then yields semi-analytic light curves with errors of $\epsilon \lesssim 0.3$ for $0.25 \leq R \leq 2$, rising to $\epsilon \lesssim 0.8$ for $R = 6$.

The results here strictly apply only to an ellipsoidal geometry. However, since the light curves arise as an integration of the emission over the entire ejecta, they depend primarily on the global morphology and only weakly on small scale distortions. The orientation effects of general, compact, contiguous geometries can thus be roughly estimated by considering a closely fitting ellipsoid with an effective axial ratio $R_\mathrm{eff}$. In what follows, though, we analyze other geometries using the geometry-specific parallel projected area method, since this approach turns out to be fairly robust.

\subsection{Torus}
\label{subsec:results_torus}

The light curves of a torus ejecta are analogous to those of an ellipsoid ejecta. In particular, we can explain the inclination-variation of the light curves using the same projected surface area method, with two main geometry-specific differences. The first is that the torus has a more complicated projected area (Section \ref{subsec:projected_area_torus}). In particular, it has a characteristic break at $\pm \mu_\mathrm{break} = \pm 1/K$, which corresponds to the viewing angle at which the torus begins to fully block the central hole. For $|\mu| \geq K^{-1/2}$, the torus does not block the hole; for $K^{-1} < |\mu| < K^{-1/2}$, it partially blocks the hole; and for $|\mu| \leq K^{-1}$, it fully blocks the hole. The projected area thus transitions from a shallower to a steeper dependence on viewing angle at $\mu = K^{-1}$ (Figure \ref{fig:projected_area}). The scale of the variation is again given by the pole-to-equator projected area ratio
\begin{equation}
\frac{A_\mathrm{proj}(\mu=1;K)}{A_\mathrm{proj}(\mu=0;K)} = \frac{4\pi K}{4K + \pi}
\end{equation}
The second is that the torus has a different reference direction $\mu_\mathrm{ref} = 0.45$. The relation $L(\tau,\mu_\textrm{ref};K) \simeq L_\mathrm{av}(\tau;K)$ has errors $\epsilon \lesssim 0.1$ with this $\mu_\mathrm{ref}$ for both the broken power-law and constant density profiles. As before, this suggests that $\mu_\mathrm{ref}$ is a function primarily of the geometry.

\begin{figure}
\includegraphics[width=0.46\textwidth]{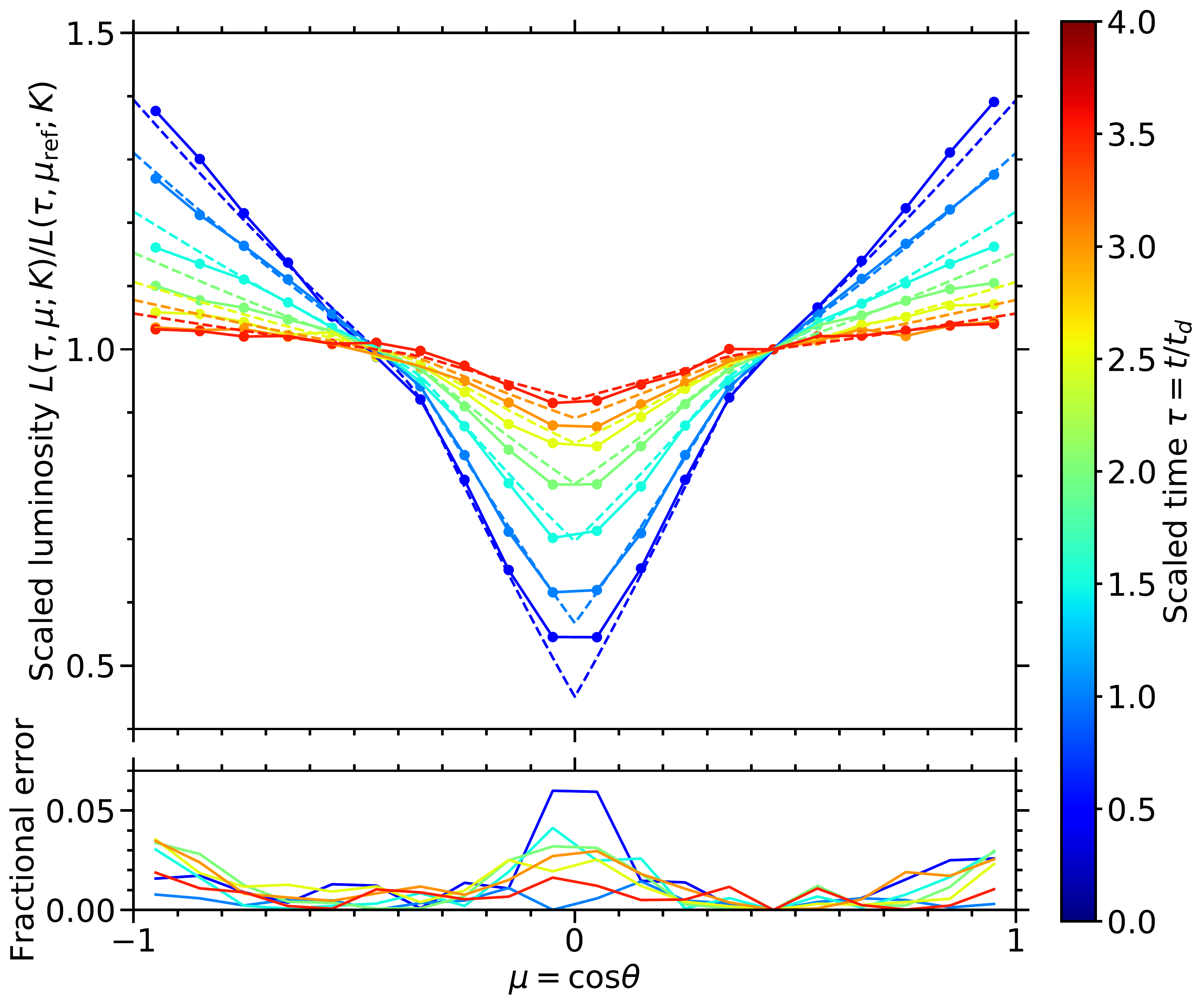}
\caption{
The isotropic-equivalent bolometric luminosities for a torus outflow with a constant density profile and radius ratio $K = 3$ as a function of viewing angle $\mu = \cos \theta$ and at different times. The time $t$ has been scaled by the diffusion time $t_d$ as $\tau = t / t_d$ (Section \ref{subsec:methods_general}), and the isotropic-equivalent luminosity $L(\tau,\mu;R)$ has been scaled by that in the reference direction $\mu_\mathrm{ref} = 0.45$, since in this direction the luminosity roughly equals the angle-averaged luminosity. The torus has a constant density profile. The thermalization time is $\tau_e = 10$. The points and solid colored curves show the luminosities at different values of $\tau < 4$. The dashed colored curves show the fits to the function in Equation \ref{eq:L_of_tau_mu_R_analytic} with the replacement $R \rightarrow K$. 
% The anomalous behavior of the curves at the earliest time are due to an initial transient before the transport converges. 
The bottom panel shows the error of the fits.
}
\label{fig:psi_of_tau_mu_vs_mu-torus}
\end{figure}

To illustrate these differences, we present the signatures for a torus ejecta with a constant density profile and a radius ratio $K = 3$. Figure \ref{fig:scaled_lightcurves-torus-constant_density} shows the isotropic-equivalent light curves. Figure \ref{fig:psi_of_tau_mu_vs_mu-torus} shows the variation of the luminosity ratio $L(\tau,\mu;K) / L(\tau,\mu_\mathrm{ref};K)$ with viewing angle at different times.

\begin{figure*}
\gridline{
\fig{{L_of_tau_mu_vs_tau-with_curve_fits-g_1-K_3-constant_density}.pdf}{0.46\textwidth}{(a) Scaled by $L_n = M q_0 t_{d,\mathrm{day}}^{\sigma_1}$}
\fig{{h_of_tau_mu_vs_tau-K_3-constant_density}.pdf}{0.46\textwidth}{(b) Scaled by $L_\mathrm{av}(\tau;K)$}
}
\caption{
The isotropic-equivalent bolometric light curves $L(\tau,\mu;K)$ for a torus outflow with a constant density profile and radius ratio $K = 3$ at different viewing angles $\mu = \cos\theta$. The time $t$ has been scaled by the diffusion time $t_d$ as $\tau = t / t_d$, and the thermalization time is $\tau_e = 10$. (a) The isotropic-equivalent luminosity has been scaled by the factor $L_n = M q_0 t_{d,\mathrm{day}}^{\sigma_1}$ (Section \ref{subsec:methods_general}). The solid black curve shows the angle-averaged luminosity, the dashed black curve shows the luminosity of the equivalent spherical ejecta (with the same $M$, $E_k$, and $\tau_e$), and the dotted black curve shows the total heating rate. The dashed colored curves show the curve fits using the functions in Equations \ref{eq:L_of_tau_mu_R_analytic} and \ref{eq:fitting_function_k}, with the replacement $R \rightarrow K$. The bottom panel shows the error of the fits. (b) The isotropic-equivalent luminosity has been scaled by the angle-averaged luminosity $L_\mathrm{av}(\tau;K)$. The solid black horizontal curve shows the angle-averaged luminosity. The bottom panel shows the error of $L(\tau,\mu_\mathrm{ref};K)$ from $L_\mathrm{av}(\tau;K)$ for $\mu_\mathrm{ref} = 0.45$.
}
\label{fig:scaled_lightcurves-torus-constant_density}
\end{figure*}

The features shown in this specific case generalize to both density profiles (broken power-law and constant) and over our range of $K \in [1,5]$, in a way analogous to the ellipsoid ejecta. The light curves in the broken power-law case have a rise-peak-fall time evolution. The relation $L_\mathrm{av}(\tau;K) \simeq L_s(\tau)$ for the constant density case has errors $\epsilon \lesssim 0.2$ for $1 \leq K \leq 2$ rising to $\epsilon \lesssim 2.2$ for $K = 5$, and for the broken power-law case has errors $\epsilon \lesssim 0.3$ for $1 \leq K \leq 2$ rising to $\epsilon \lesssim 3.7$ for $K = 5$. The fitting parameters $k(\tau;K)$ have similar shapes but different values compared to the ellipsoidal case, and thus the analytic approximation for $k(\tau;K)$ has different values for $k_0(K)$ and $\tau_b(K)$ (Figure \ref{fig:k_vs_tau-torus}). It is interesting to compare the values for $K \in [1,5]$ to those for $R \in [2,6]$ since they have similar degrees of asphericity; in general, the parameters $k_0$ and $\tau_b$ appear to depend on both the geometry and density profile. The semi-analytic light curves for the broken power-law case have errors $\epsilon < 0.4$ for $1 \leq K \leq 2$ rising to $\epsilon < 0.8$ for $K = 5$, and those for the constant density case have errors $\epsilon < 0.55$ for $1 \leq K \leq 2$ rising to $\epsilon < 0.75$ for $K = 5$.

\section{Discussion}
\label{sec:discussion}

The light curves presented here used ejecta models with idealized geometries and constant opacity to capture the effects of asymmetry. The idealized approach provides intuition for the fundamentally geometric effect that global asymmetry has on the light curves. Given that approximate models are frequently used to estimate kilonova light curves, the synthetic light curves presented here and the simple analytic prescriptions for them (Equations \ref{eq:L_of_tau_mu_R_analytic} and \ref{eq:fitting_function_k}) should be valuable for estimating the impact of asphericity and orientation.

We have focused here on the bolometric light curves of kilonovae and only marginally discussed their spectra and colors. However, the scale invariance of our models leads to dimensionless light curves, which we argued can serve as surrogates for the unobstructed emission produced by the different ejecta components, after reintroducing the appropriate scales of ejecta mass, opacity, and expansion velocity. Indeed, though the models used a uniform composition and a grey opacity, the resulting blackbody spectra provide a workable approximation of the colors. The colors are sensitive to the composition of the ejecta, tending to redder wavelengths when high-opacity lanthanide species are present \citep{kasen13}. 
% For ejecta with uniform composition, models with grey opacity that produce blackbody spectra provide a workable approximation of the colors.

In general, though, kilonova ejecta have multiple interacting components, compositional inhomogeneities, and multiwavelength opacity. This structure implies that blocking, reprocessing, and funneling effects, in addition to geometric effects, may also produce substantial color variation with viewing angle \citep{perego14,perego17,kasen15,kasen17,wollaeger18,kawaguchi18,kawaguchi19,bulla19}. The extent and efficacy of these depend on the spatial distribution and properties of the low-$Y_e$ (lanthanide-rich, high opacity) material of dynamical origin and low- to high-$Y_e$ (lanthanide-poor, low opacity) material of wind origin. 
For example, current dynamical simulations show that the lanthanide-rich material concentrates torus-like in the equatorial region and the lanthanide-poor material concentrates cone-like in the polar regions (e.g. \citealt{radice18a}). If the lanthanide-rich material has a larger radius and expansion velocity than the lanthanide-poor material, then the former can modify the blue emission from the latter: blocking inhibits the emission from penetrating along the equator \citep{kasen15,kasen17,wollaeger18,kawaguchi18,kawaguchi19,bulla19}; reprocessing shifts the observed photosphere to larger radii and velocities, reduces the effective temperatures and frequencies, and isotropizes the direction \citep{perego17,kawaguchi18,kawaguchi19,bulla19}; and funneling redirects the diffusing photons in the polar direction \citep{kawaguchi19}. The aggregate result is that the blue emission is obscured in the equatorial direction, and thus bluer spectra are observed at near-polar angles and redder spectra at near-equatorial angles. In contrast, if the lanthanide-rich material is enshrouded in lanthanide-poor material, then analogously, the red emission from the former is reprocessed into blue emission and isotropized \citep{kawaguchi19}.

More simply, we can use the projected surface area approach to make a rough estimate of the impact of geometry and blocking from this ejecta. We model the outflow as a cone of low-opacity ejecta with half-opening angle $\theta_c \in [0,\pi/2]$ embedded inside a high-opacity sphere of radius $a$ (Figure \ref{fig:geometries}). In this model, only the  polar ``caps'' of the low-opacity cone will be visible, as the interior is obscured by the high-opacity lanthanide-rich envelope. The projected surface area $A_\mathrm{proj}^\mathrm{caps}(\mu;\mu_c)$ of the caps then provide a rough estimate of the viewing angle dependence of the blue light curve, while that for the red light curve can be described by the complementary geometry $A_\mathrm{proj}^\mathrm{comp}(\mu;\mu_c) = \pi a^2 - A_\mathrm{proj}^\mathrm{caps}(\mu;\mu_c)$.

Appendix \ref{subsec:projected_area_conical_cap} gives analytic formulae for the projected surface area of the conical caps, and Figure~\ref{fig:projected_area} shows the variation of the area with viewing angle. The results indicate that the blue light curve will be brighter for polar viewing angles where the cap is fully visible, and dimmer for equatorial ones where only a small region of the cap is visible. The pole-to-equator projected area ratio has the simple analytic expression
\begin{equation}
\frac{A_\mathrm{proj}(\mu=1;\mu_c)}{A_\mathrm{proj}(\mu=0;\mu_c)} = \frac{\pi \sin^2\theta_c}{2 ( \theta_c - \sin\theta_c \cos\theta_c )}
\end{equation}
For a typical conical opening angle $\theta_c \approx 45^\circ$ ($\mu_c \approx 0.71$) \citep{radice18a}, the projected area ratio is $\approx 2.8$. We thus expect the blue luminosity to vary by a factor of $\sim 3 - 5$ with viewing angle at peak, consistent with prior detailed transport models \citep{kasen15,wollaeger18,kawaguchi18,kawaguchi19,bulla19}. For smaller opening angles $\theta_c \approx 20^\circ$ ($\mu_c \approx 0.94$), the variation can be a factor of $\sim 5 - 10$ at peak, i.e. up to an order of magnitude. While detailed calculations are needed for quantitative modeling, this semi-analytic approach using the projected surface area roughly gauges the size of the geometric effects. A similar approach could presumably be used for modified geometries as well. For instance, this conical cap model assumes the lanthanide-poor and lanthanide-rich material have a comparable radial extant, whereas a modified setup could potentially accommodate two different component velocities.

\begin{figure*}
\gridline{
\fig{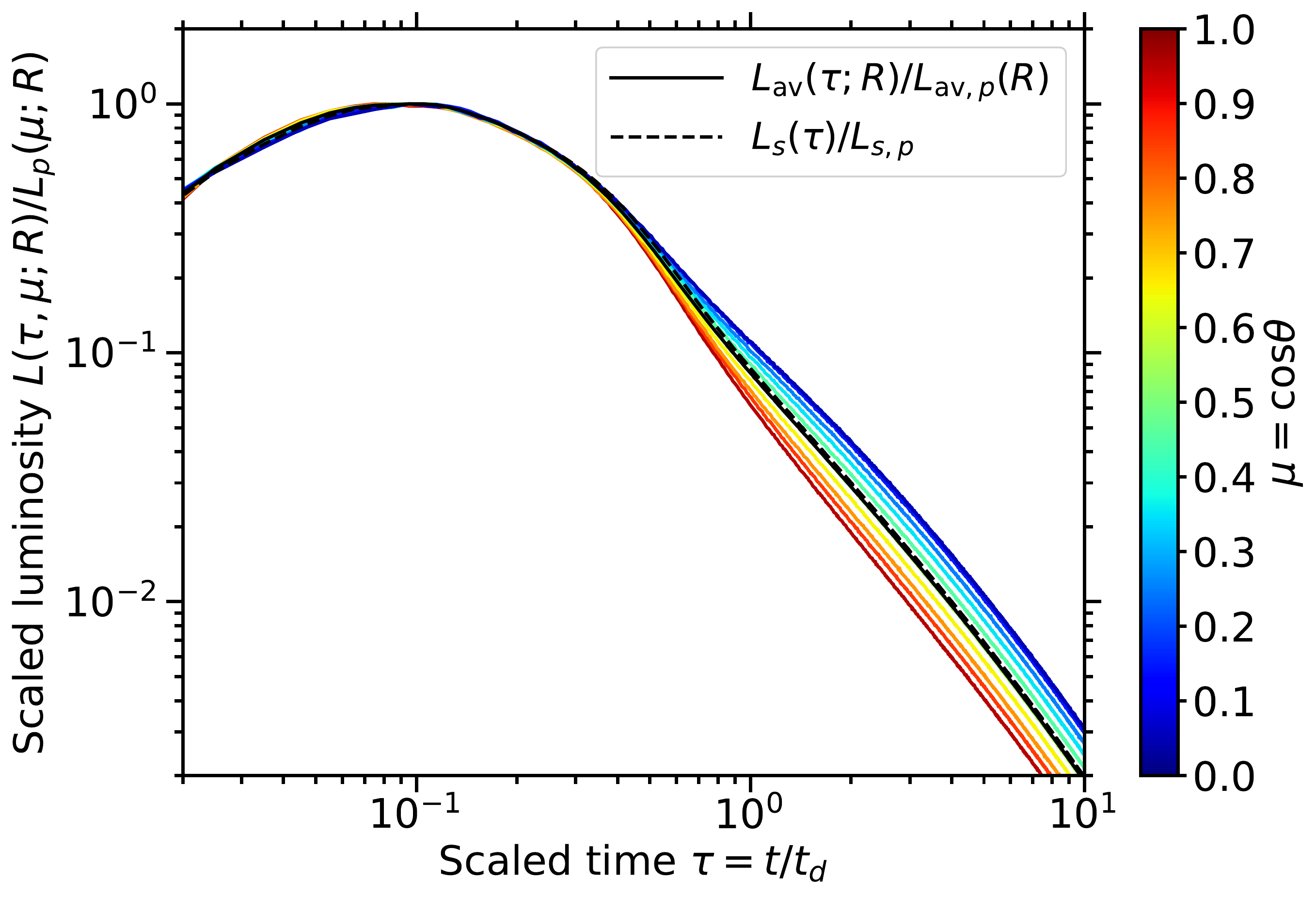}{0.46\textwidth}{(a) $R = 2$, varying $\mu$}
\fig{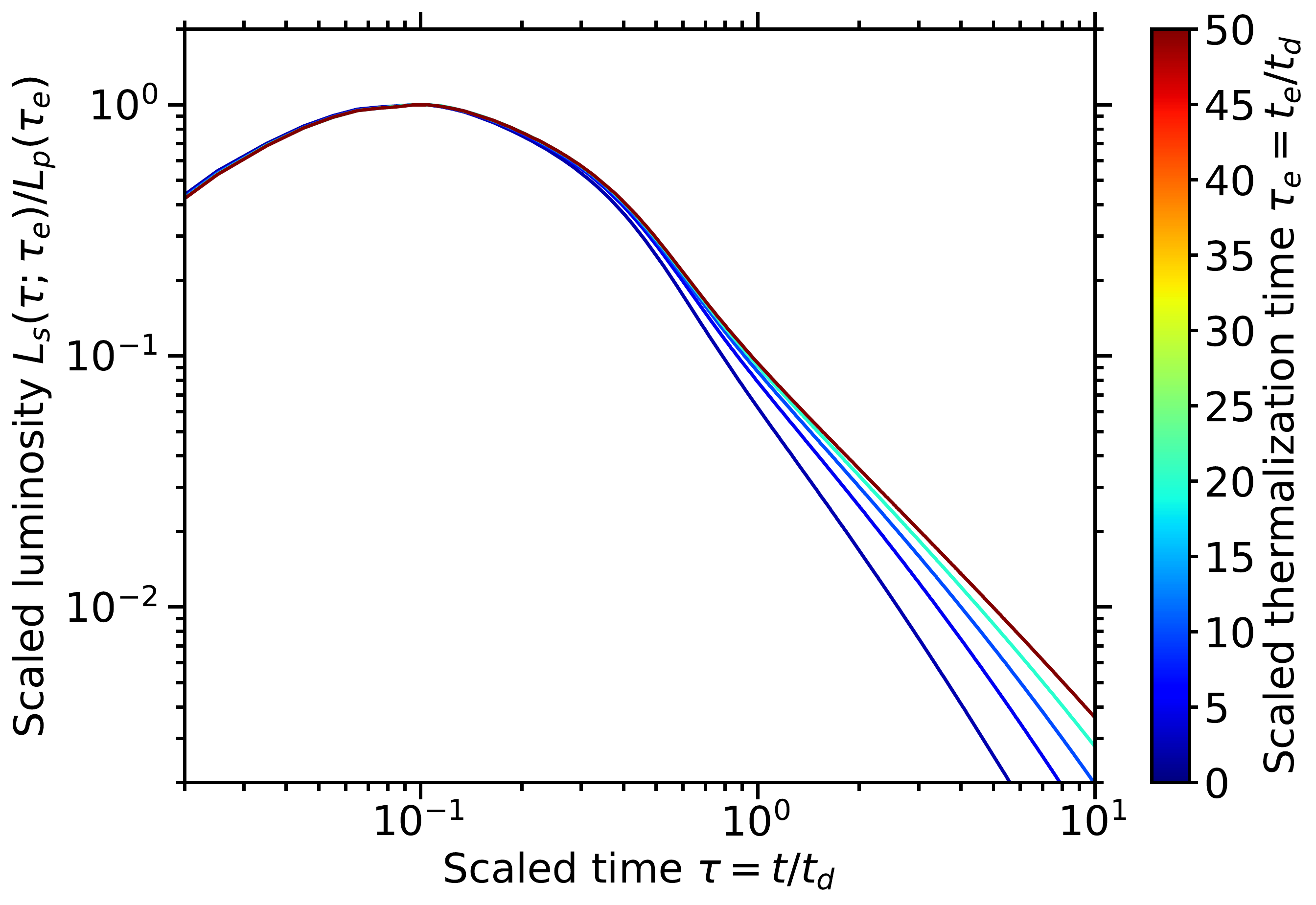}{0.46\textwidth}{(b) $R = 1$, varying $\tau_e$}
}
\caption{(a) The isotropic-equivalent bolometric light curves for an ellipsoid outflow with axial ratio $R = 2$ at different viewing angles $\mu = \cos \theta$. The time $t$ has been scaled by the diffusion time $t_d$ as $\tau = t / t_d$ (Section \ref{subsec:methods_general}), and the isotropic equivalent luminosity $L(\tau,\mu;R)$ has been scaled by its peak value $L_p(\mu;R)$. The solid black curve shows the angle-averaged luminosity, and the dashed black curve shows the luminosity of the equivalent spherical ejecta ($R=1$ with the same $\tau_e$). (b) The bolometric light curve of a spherical outflow with different thermalization times $\tau_e = t_e / t_d$. The luminosity has been scaled by its peak value $L_p(\tau_e)$. We note that $L_p(\tau_e)$ varies little for the values of $\tau_e$ shown since they satisfy $\tau_e \gg \tau_p(\mu;R) \sim 0.1$. A comparison of the two panels shows that there is some degeneracy between observing 1) an ellipsoidal outflow at different viewing angles and 2) a spherical outflow with different heating rates, though the two may be distinguishable by careful measurement of the power-law decline at late times.}
\label{fig:degenerate_light curves}
\end{figure*}

The results in this paper suggest potential diagnostics for inferring the orientation angle fron kilonova observations. As the orientation of a merging binary system is only partially constrained by its GW signal, a complementary constraint from kilonova observations would be of considerable interest. Since the inclination-dependence of the light curves diminishes with time, the ratio of the luminosity on the light curve tail to that at peak is highly correlated with orientation, as shown in Figure \ref{fig:degenerate_light curves}. The tail-to-peak ratio declines more rapidly for viewing angles with larger projected areas since these have higher peaks and must converge to the total heating rate $Q(\tau;\tau_e)$ at late times. The ratio of curves of different $\mu$ at late times equals the ratio of their peak luminosities $L_p(\mu;R)$. Ejecta with higher $R$ have a larger spread in decline rates since the projected area changes more from pole to equator.

The tail-to-peak luminosity ratio thus provides a potential way to constrain the outflow geometry and inclination. However, the luminosity ratio is partly degenerate with other inputs, such as the heating rate (Figure \ref{fig:degenerate_light curves}), which complicates efforts to use the light curve alone to infer the ejecta geometry or the underlying radioactivity and thermalization. It remains to be seen the extent to which  multi-parameter fitting of the light curves can individually constrain the parameters.

The parameters (mass, velocity, etc.) of the blue and red components of GW170817/AT2017gfo were inferred by fitting the observed light curves with synthetic ones from spherically symmetric models \citep{cowperthwaite17,chornock17,nicholl17,kasen17,tanaka17}. Aspherical models would generally lead to different inferred parameters. To quantify the size of the difference, we compare a spherical model (with parameters labeled with the subscript $s$) to ellipsoidal models with axial ratios $R = 1/2$ and $2$, which encapsulate the two directions the asphericity might tend. Indeed, the red component likely comes from the post-merger disk wind, which is mildly prolate, but the source of the blue component is more uncertain, and could arise from the dynamical ejecta or more hypothetical outflows \citep{metzger17}. We use the current estimates for the viewing angle, $\theta \approx 30^\circ$ ($\mu \approx 0.87$) \citep{finstad18,abbott19,wu19}. 
% The procedure is simple. We assume that both the spherical and ellipsoidal models have $\tau_e \gg \tau_p(\mu;R) \sim 0.1$ so that their dimensionless luminosities at peak, $\Lambda_{s,p}$ and $\Lambda_p$, do not change appreciably with $\tau_e$. For $R = 1/2$, the dimensionless peak luminosity is $\Lambda_p \sim 0.7 \Lambda_{s,p}$. For $R = 2$, it is $\Lambda_p \sim 1.7 \Lambda_{s,p}$. We assume that $t_d$ is the same for both the spherical and ellipsoidal models so that they have the same physical time to peak $t_p(\mu;R)$, and that $q_0$, $\sigma_1$, and $\sigma_2$ are the same as well. We then reintroduce dimensions, and match the peaks ($L_{s,p}$ and $L_p$) and late-time light curves (at $\tau \gg \tau_e$) to infer the parameters. For $R=1/2$, we find $M \sim 1.4 M_s$, $\kappa/\beta_\mathrm{ch} \sim 0.7 (\kappa/\beta_\mathrm{ch})_s$, and $\tau_e \sim 0.8 (\tau_e)_s$. For $R=2$, we find $M \sim 0.6 M_s$, $\kappa/\beta_\mathrm{ch} \sim 1.7 (\kappa/\beta_\mathrm{ch})_s$, and $\tau_e \sim 1.5 (\tau_e)_s$. 
We then infer the mass, opacity, velocity, and thermalization time that yield the same peak and late-time behavior in the physical light curve as in the spherical case. For $R=1/2$, we find $M \sim 1.4 M_s$ (i.e. $\sim 40 \%$ larger than the spherical case), $\kappa/\beta_\mathrm{ch} \sim 0.7 (\kappa/\beta_\mathrm{ch})_s$, and $\tau_e \sim 0.8 (\tau_e)_s$. For $R=2$, we find $M \sim 0.6 M_s$ (i.e. $\sim 40 \%$ smaller than the spherical case), $\kappa/\beta_\mathrm{ch} \sim 1.7 (\kappa/\beta_\mathrm{ch})_s$, and $\tau_e \sim 1.5 (\tau_e)_s$. The mass uncertainties from geometric effects of this magnitude are a factor $\lesssim 2$, and are comparable to those from uncertainties in the ejecta opacities \citep{barnes13} and the heating rate \citep{metzger10,korobkin12,lippuner15}.

In BH-NS mergers, the tidal component of the dynamical ejecta can be highly asymmetric. In general terms, if the binary has a low mass ratio, the BH has high aligned spin, and the NS has low compactness, then tidal disruption can occur far outside the BH horizon, leading to a single tidal tail concentrated in the equatorial plane $(v_x / v_z \sim 5)$ with a limited azimuthal range $(\Delta \phi \lesssim \pi)$ \citep{foucart13,foucart19,kyutoku15}. Though this is a typical case, the detailed properties of the tidal tail and the fraction of unbound mass depend sensitively on the particulars of the merger, such as the BH spin alignment and the NS equation of state \citep{kyutoku15,kawaguchi15,foucart17}. We can roughly model the tidal tail as an ellipsoid with axial ratio $R = 5$ or a torus with radius ratio $K = 5$. The luminosity variation with polar viewing angle is then roughly a factor of $\sim 5 - 10$ near peak, before reducing to $\sim 1$ at late times. Global geometric effects are thus important inputs to accurately constrain the mass ejected from these systems.

%% If you wish to include an acknowledgments section in your paper,
%% separate it off from the body of the text using the \acknowledgments
%% command.
\acknowledgements
{
We thank the anonymous reviewer for helpful comments. This research used resources of the National Energy Research Scientific Computing Center, a Department of Energy Office of Science User Facility supported by the Office of Science of the U.S. Department of Energy under Contract No. DE-AC02-05CH11231. This work was supported in part by the U.S. Department of Energy, Office of Science, Office of Nuclear Physics, under contract number DE-AC02-05CH11231 and DE-SC0017616, by a SciDAC award DE-SC0018297, by the Gordon and Betty Moore Foundation through Grant GBMF5076, and by the National Science Foundation under Grant No. 1616754.
}

%% To help institutions obtain information on the effectiveness of their 
%% telescopes the AAS Journals has created a group of keywords for telescope 
%% facilities.
%
%% Following the acknowledgments section, use the following syntax and the
%% \facility{} or \facilities{} macros to list the keywords of facilities used 
%% in the research for the paper.  Each keyword is check against the master 
%% list during copy editing.  Individual instruments can be provided in 
%% parentheses, after the keyword, but they are not verified.

% \vspace{5mm}
% \facilities{HST(STIS), Swift(XRT and UVOT), AAVSO, CTIO:1.3m,
% CTIO:1.5m,CXO}

%% Similar to \facility{}, there is the optional \software command to allow 
%% authors a place to specify which programs were used during the creation of 
%% the manusscript. Authors should list each code and include either a
%% citation or url to the code inside ()s when available.

\software{
SEDONA \citep{kasen06}
}

%% Appendix material should be preceded with a single \appendix command.
%% There should be a \section command for each appendix. Mark appendix
%% subsections with the same markup you use in the main body of the paper.

%% Each Appendix (indicated with \section) will be lettered A, B, C, etc.
%% The equation counter will reset when it encounters the \appendix
%% command and will number appendix equations (A1), (A2), etc. The
%% Figure and Table counter will not reset.

\appendix

\section{Projected Area}
\label{sec:projected_area}

\begin{figure*}
\gridline{
\fig{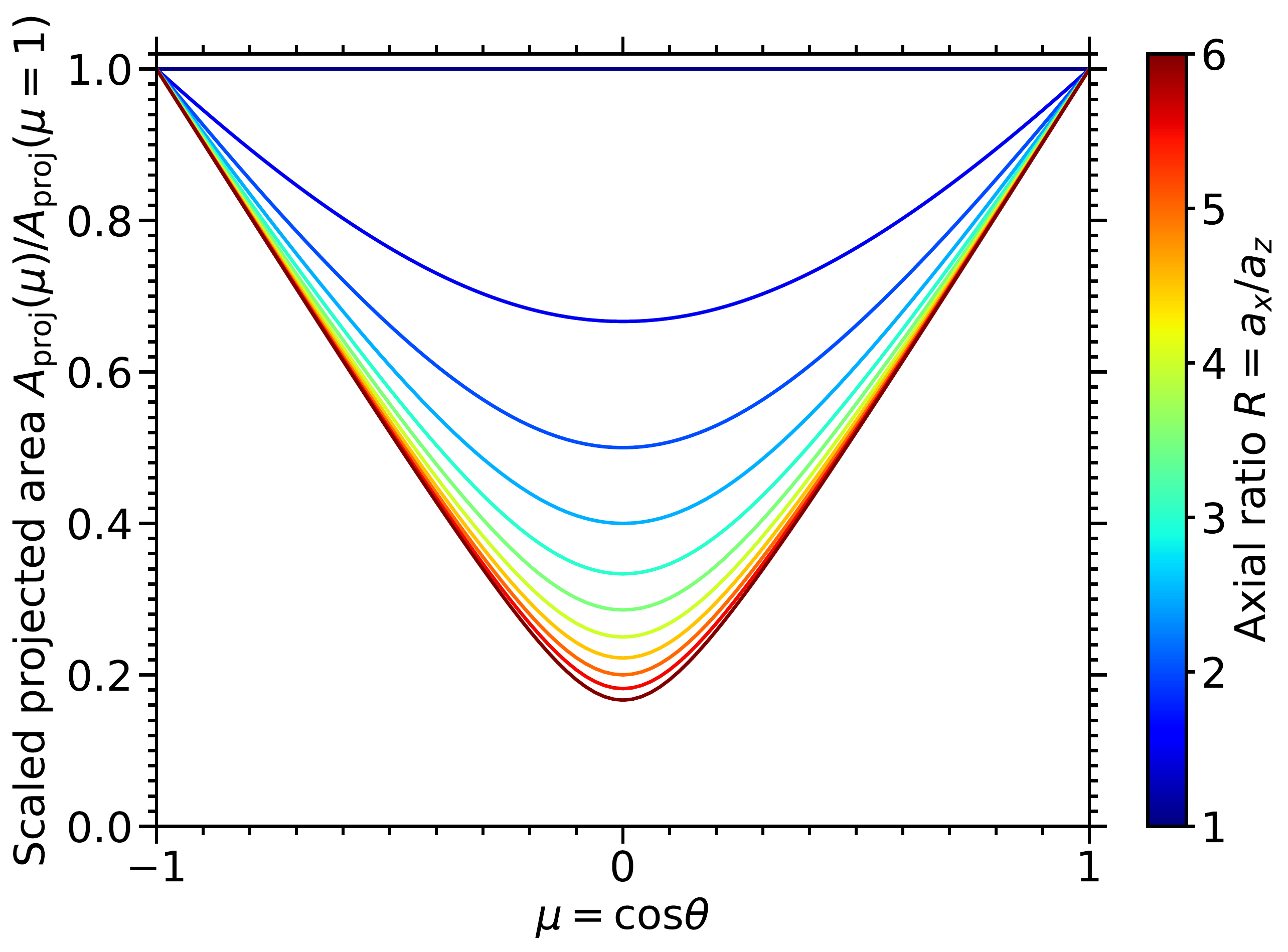}{0.32\textwidth}{(a) Ellipsoid ($R>1$)}
\fig{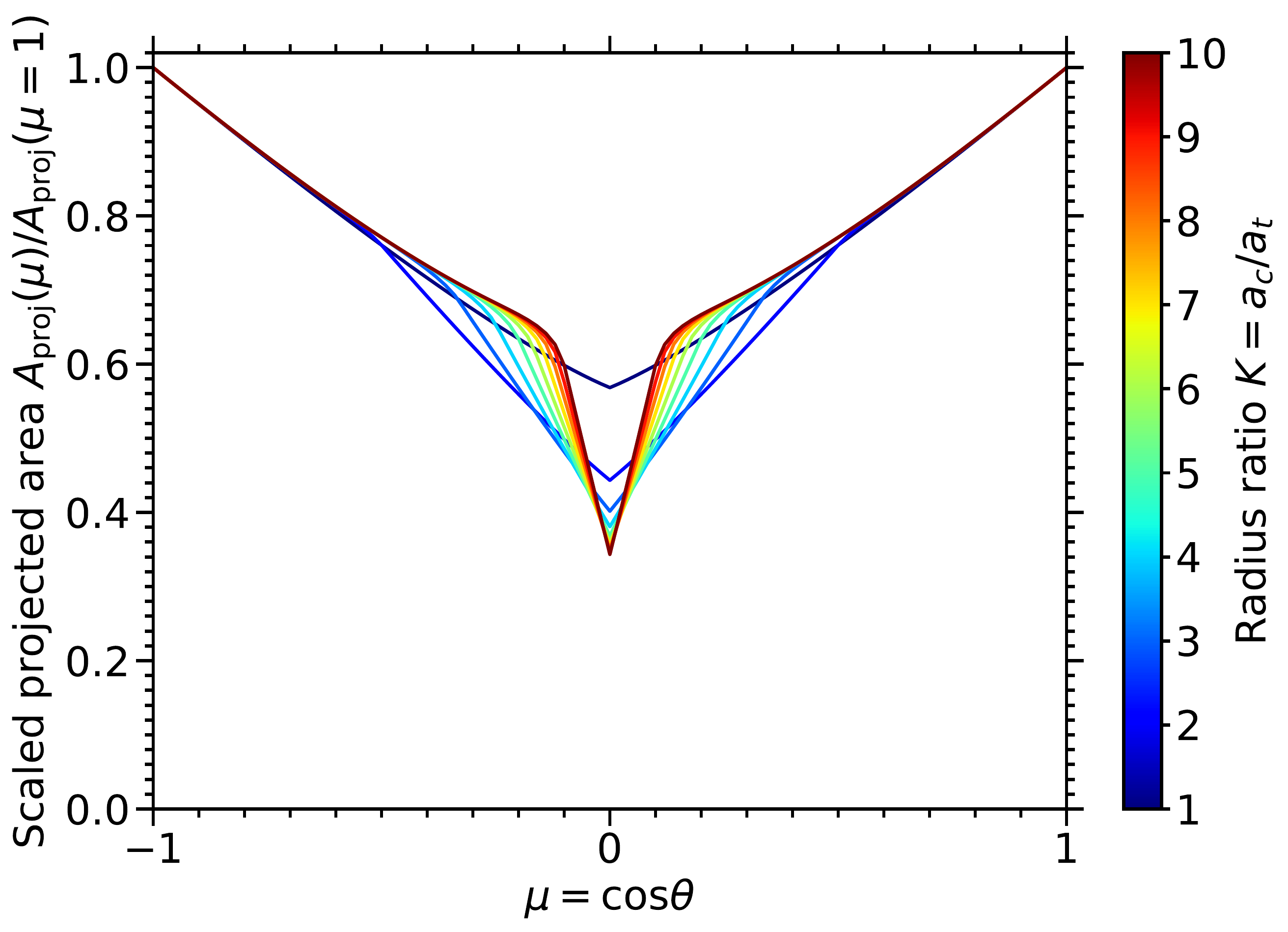}{0.32\textwidth}{(b) Torus ($K>1$)}
\fig{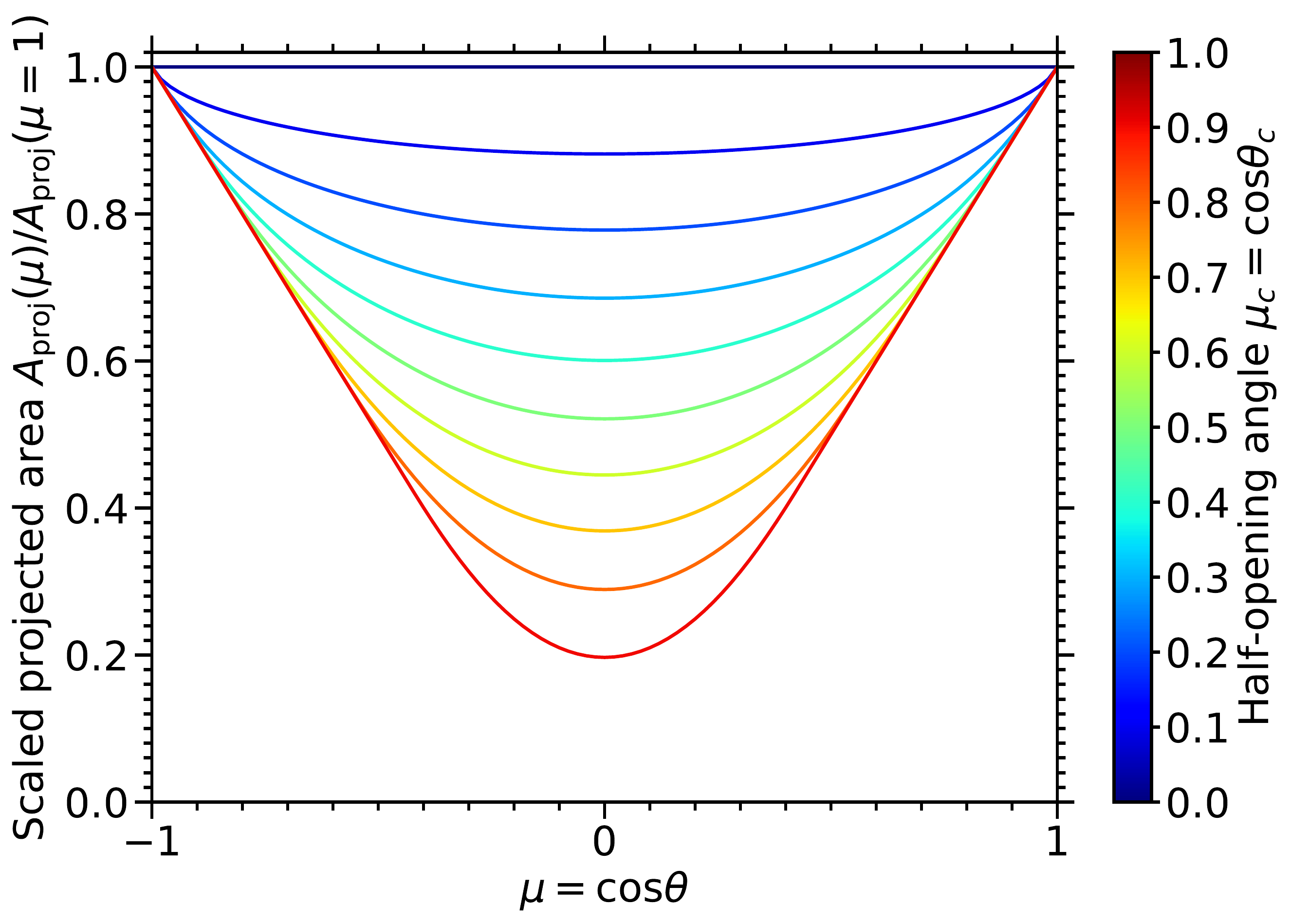}{0.32\textwidth}{(c) Conical cap}
}
\caption{The analytic projected area $A_\mathrm{proj}(\mu;R)$ for several axisymmetric geometries. The panels show the projected area for (a) an ellipsoid with axial ratio $R = a_x/a_z > 1$, (b) a ring torus with radius ratio $K = a_c/a_t > 1$, and (c) two conical caps with half-opening angle $\mu_c \in [0,1]$.  Equations for the analytic functions $A_{\rm proj}(\mu)$ are given in the text (Appendix \ref{sec:projected_area}).
}
\label{fig:projected_area}
\end{figure*}

In this appendix, we present equations for the parallel projected areas of the ejecta geometries studied in the paper (Figure \ref{fig:geometries}): an ellipsoid, a ring torus, and a conical cap. The geometries are axisymmetric, so we parameterize the viewing angle with $\mu = \cos\theta$, where $\theta$ is the polar angle. Figure \ref{fig:projected_area} shows the projected areas as a function of $\mu$, normalized to their values at $\mu = \pm 1$.

\subsection{Ellipsoid}
\label{subsec:projected_area_ellipsoid}

We study an ellipsoid with semi-major axes $(a_x,a_y,a_z)$, where $a_x = a_y$ (axisymmetric, spheroid), and we define the axial ratio $R = a_x/a_z$ (Figure \ref{fig:geometries}). A straightforward analysis of the geometry shows that the parallel projected area of the ellipsoid in the direction $\mu$ is an ellipse with semi-major axes $(a_x, a_z [ (R^2 - 1) \mu^2 + 1 ]^{1/2})$. The parallel projected area in the direction $\mu$ is thus
\begin{equation}
A_\mathrm{proj}(\mu;R) = \pi R a_z^2 [ (R^2 - 1) \mu^2 + 1 ]^{1/2}
\end{equation}
Figure \ref{fig:projected_area} shows a plot of $A_\mathrm{proj}(\mu;R)$. The pole-to-equator ratio is
\begin{equation}
\frac{A_\mathrm{proj}(\mu=1;R)}{A_\mathrm{proj}(\mu=0;R)} = R
\end{equation}

\subsection{Torus}
\label{subsec:projected_area_torus}

We study a torus with spine radius $a_c$ and tube radius $a_t$, where $a_c > a_t$ (ring torus), and we define the radius ratio $K = a_c/a_t \geq 1$ (Figure \ref{fig:geometries}). 
% The parallel projected area can be found by considering one method of constructing a torus. To construct a torus, one can take a circular spine of radius $a_c$, place a sphere of radius $a_t$ with its center on the spine, and revolve the sphere along the spine. The projected area of the circular spine in the direction $\mu$ is an ellipse with semi-major axes $(a_c, a_c \mu)$, and the projected area of the sphere is a circle of radius $a_t$ with its center on the spine. To construct the projected area of the torus, one can thus take an ellipse with semi-major axes $(a_c, a_c \mu)$, place a circle of radius $a_t$ with its center on the ellipse, and revolve the circle along the ellipse, i.e. the projected area is simply the area bounded by the parallel curves of an ellipse with semi-major axes $(a_c, a_c \mu)$. 
A straightforward analysis of the geometry shows that the parallel projected area of the torus in the direction $\mu$ is the area bounded by the parallel curves of an ellipse with semi-major axes $(a_c, a_c \mu)$. The outer $(+)$ and inner $(-)$ parallel curves can be expressed with the parametric equations
\begin{align}
\tilde{x}(t) &= a_t \left( K \pm \frac{\mu}{(1 - (1-\mu^2) \cos^2 t)^{1/2}} \right) \cos t \\
\tilde{y}(t) &= a_t \left( K \cos\theta \pm \frac{1}{(1 - (1-\mu^2) \cos^2 t)^{1/2}} \right) \sin t
\end{align}

The parallel projected area in the direction $\mu$ can be written as
\begin{equation}
A_\mathrm{proj}(\mu;K) = A_\mathrm{outer}(\mu;K) - A_\mathrm{inner}(\mu;K)
\end{equation}
where $A_\mathrm{outer}(\mu;K)$ and $A_\mathrm{inner}(\mu;K)$ are the areas contained inside the outer and inner parallel curves, respectively. The following integrals will be useful for evaluating these areas:
\begin{align}
I_1(t_1,t_2) &= \int_{t_1}^{t_2} \sin^2 t dt \\
I_2(\mu;t_1,t_2) &= \int_{t_1}^{t_2} \frac{\sin^2 t}{(1 - (1-\mu^2) \cos^2 t)^{3/2}} dt \\
I_3(\mu;t_1,t_2) &= \int_{t_1}^{t_2} \frac{\sin^2 t}{(1 - (1-\mu^2) \cos^2 t)^{1/2}} dt \\
I_4(\mu;t_1,t_2) &= \int_{t_1}^{t_2} \frac{\sin^2 t}{(1 - (1-\mu^2) \cos^2 t)^2} dt
\end{align}
We note that $I_1(0,\pi) = \pi/2$ and $I_4(\mu;0,\pi) = \pi/2|\mu|$. The outer projected area can be written as
\begin{equation}
A_\mathrm{outer}(\mu;K) = 2 a_t^2 \left[ K^2 |\mu| I_1(0,\pi) + K \mu^2 I_2(\mu;0,\pi) + K I_3(\mu;0,\pi) + |\mu| I_4(\mu;0,\pi) \right]
\end{equation}
The form of the inner projected area depends on the range of $\mu$: for different $\mu$, the torus blocks the central hole to a different degree, which leads to a different degree of overlap of the inner parallel curve with itself. The inner projected area can be divided into three domains
\begin{enumerate}

% \item $K|\mu| \geq K^{1/2}$: no blocking/overlap
\item $|\mu| \geq K^{-1/2}$: no blocking/overlap
\begin{equation}
A_\mathrm{inner}(\mu;K) = 2 a_t^2 \left[ K^2 |\mu| I_1(0,\pi) - K \mu^2 I_2(\mu;0,\pi) - K I_3(\mu;0,\pi) + |\mu| I_4(\mu;0,\pi) \right]
\end{equation}

% \item $K^{1/2} > K|\mu| > 1$: partial blocking/overlap
\item $K^{-1/2} > |\mu| > K^{-1}$: partial blocking/overlap
\begin{equation}
A_\mathrm{inner}(\mu;K) = 2 a_t^2 \left[ K^2 |\mu| I_1(t_r,\pi-t_r) - K \mu^2 I_2(\mu;t_r,\pi-t_r) - K I_3(\mu;t_r,\pi-t_r) + |\mu| I_4(\mu;t_r,\pi-t_r) \right]
\end{equation}
where $t_r(\mu;K) = \cos^{-1} \left( (1-\mu^2)^{-1/2} \left[ 1 - K^{-2} \mu^{-2} \right]^{1/2} \right)$.

% \item $1 \geq K|\mu|$: full blocking/overlap
\item $K^{-1} \geq |\mu|$: full blocking/overlap
\begin{equation}
A_\mathrm{inner}(\mu;K) = 0
\end{equation}

\end{enumerate}
All of the integrals above can be evaluated either explicitly or in terms of elliptic integrals. Figure \ref{fig:projected_area} shows a plot of $A_\mathrm{proj}(\mu;K)$. The curves show a characteristic break at $\pm \mu_\mathrm{break} = \pm 1/K$, which corresponds to the viewing angle at which the torus begins to fully block the central hole. The pole-to-equator ratio is
\begin{equation}
\frac{A_\mathrm{proj}(\mu=1;K)}{A_\mathrm{proj}(\mu=0;K)} = \frac{4\pi K}{4K + \pi}
\end{equation}

\subsection{Conical Cap}
\label{subsec:projected_area_conical_cap}

We study a cone with half-opening angle $\theta_c \in [0,\pi/2]$ ($\mu_c = \cos\theta_c \in [0,1]$) embedded inside a sphere of radius $a$ (Figure \ref{fig:geometries}). The parallel projected area has a different form depending on the viewing angle, and can be divided into four domains. In the interval $-(1 - \mu_c^2)^{1/2} < \mu < (1 - \mu_c^2)^{1/2}$, the following variables will be useful
\begin{align}
\theta_p(\mu;\mu_c) &= \cos^{-1} \left( \frac{\mu_c}{(1-\mu^2)^{1/2}} \right) \in [0,\theta_c] \\
\theta_d(\mu;\mu_c) &= \tan^{-1} \left( \frac{\sin\theta_p}{\mu_c} \frac{(1-\mu^2)^{1/2}}{|\mu|} \right) \in \left[0,\frac{\pi}{2}\right]
\end{align}

The parallel projected area of the top and bottom caps together, when the rest of the cone is blocked by the sphere, can be written as
\begin{equation}
A_\mathrm{proj}(\mu;\mu_c) = A_\mathrm{proj}^\mathrm{top}(\mu;\mu_c) + A_\mathrm{proj}^\mathrm{top}(-\mu;\mu_c)
\end{equation}
where $A_\mathrm{proj}^\mathrm{top}(\mu;\mu_c)$ is the parallel projected area of the top cap alone, when the rest of the cone is blocked by the sphere, which takes the following form in the four domains
\begin{enumerate}

\item $1 \geq \mu \geq (1 - \mu_c^2)^{1/2}$:
\begin{equation}
A_\mathrm{proj}^\mathrm{top}(\mu;\mu_c) = \pi a^2 (1 - \mu_c^2) \mu
\end{equation}

\item $(1 - \mu_c^2)^{1/2} > \mu \geq 0$:
\begin{equation}
A_\mathrm{proj}^\mathrm{top}(\mu;\mu_c) = a^2 ( \theta_p - \sin\theta_p \cos\theta_p ) - a^2 (1-\mu_c^2)^{1/2} \mu ( \theta_d - \sin\theta_d \cos\theta_d - \pi )
\end{equation}

\item $0 > \mu > -(1-\mu_c^2)^{1/2}$:
\begin{equation}
A_\mathrm{proj}^\mathrm{top}(\mu;\mu_c) = a^2 ( \theta_p - \sin\theta_p \cos\theta_p ) + a^2 (1-\mu_c^2)^{1/2} \mu ( \theta_d - \sin\theta_d \cos\theta_d )
\end{equation}

\item $-(1-\mu_c^2)^{1/2} \geq \mu \geq -1$:
\begin{equation}
A_\mathrm{proj}^\mathrm{top}(\mu;\mu_c) = 0
\end{equation}

\end{enumerate}
Figure \ref{fig:projected_area} shows a plot of $A_\mathrm{proj}(\mu;\mu_c)$. The pole-to-equator ratio is
\begin{equation}
\frac{A_\mathrm{proj}(\mu=1;\mu_c)}{A_\mathrm{proj}(\mu=0;\mu_c)} = \frac{\pi \sin^2\theta_c}{2 ( \theta_c - \sin\theta_c \cos\theta_c )}
\end{equation}

\section{Additional Features of the Light Curves}
\label{sec:additional_features}

The projection factor presented in Section \ref{sec:results} involves several approximations and parameters. We present here data on the size of the approximations and the values of the parameters, and quantify their limitations and domains of applicability. We present the data for both geometries (ellipsoid and torus) and density profiles (broken power-law and constant), but focus in the text on an ellipsoid kilonova.

\subsection{Analytic Approximation for $k$}
\label{subsec:additional_features_k}

\begin{figure*}
\gridline{
\fig{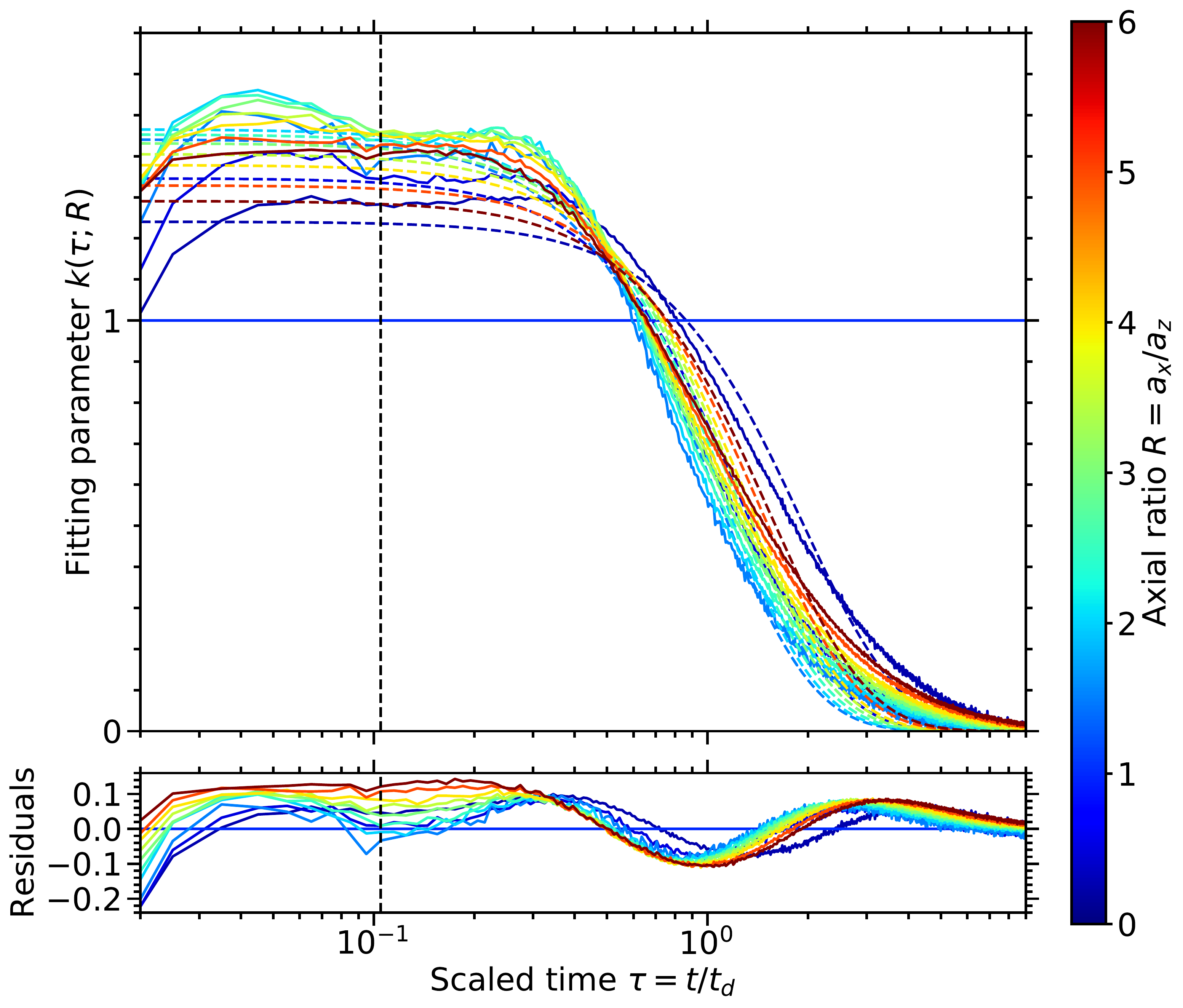}{0.32\textwidth}{(a) }
\fig{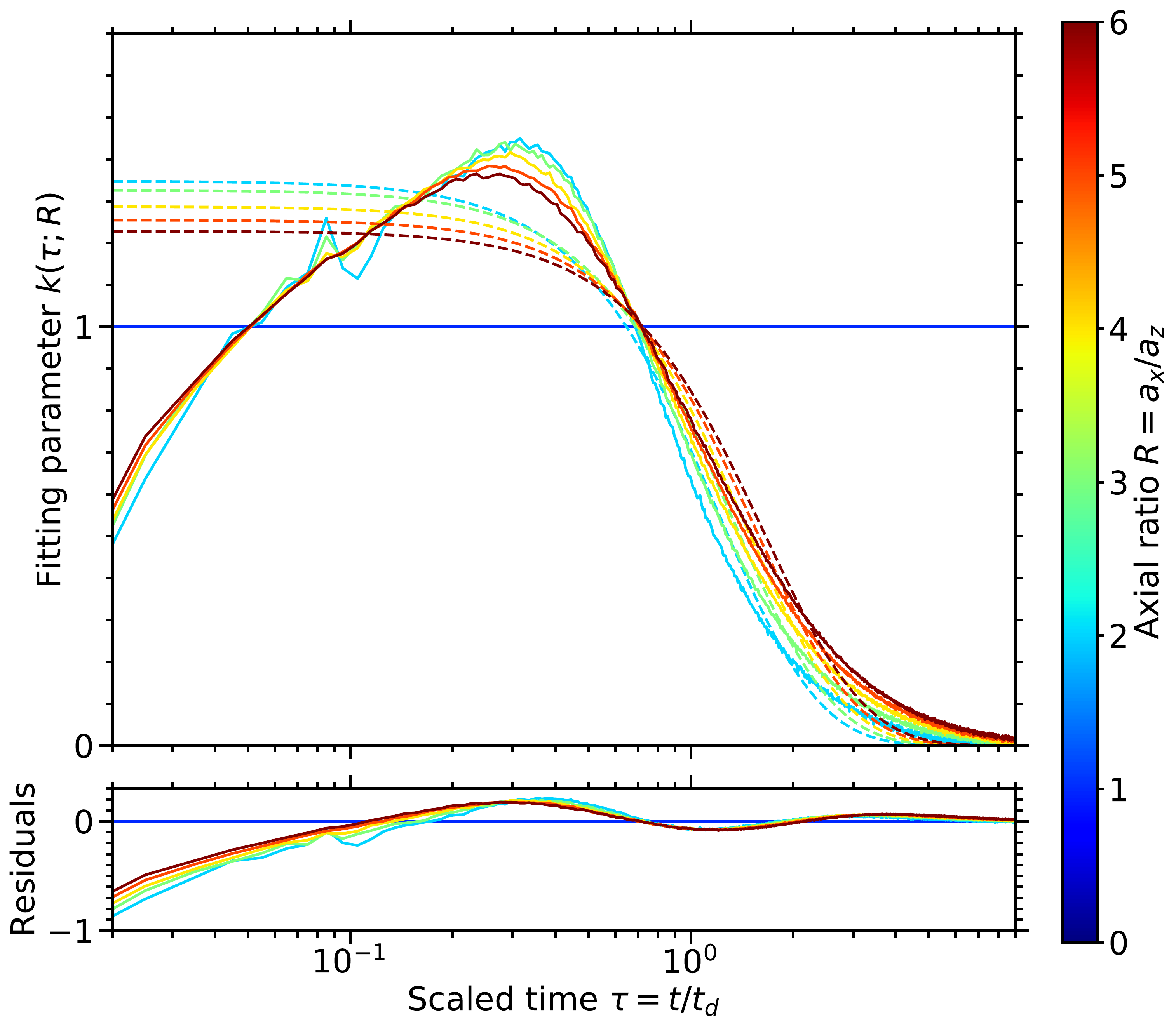}{0.32\textwidth}{(b) }
\fig{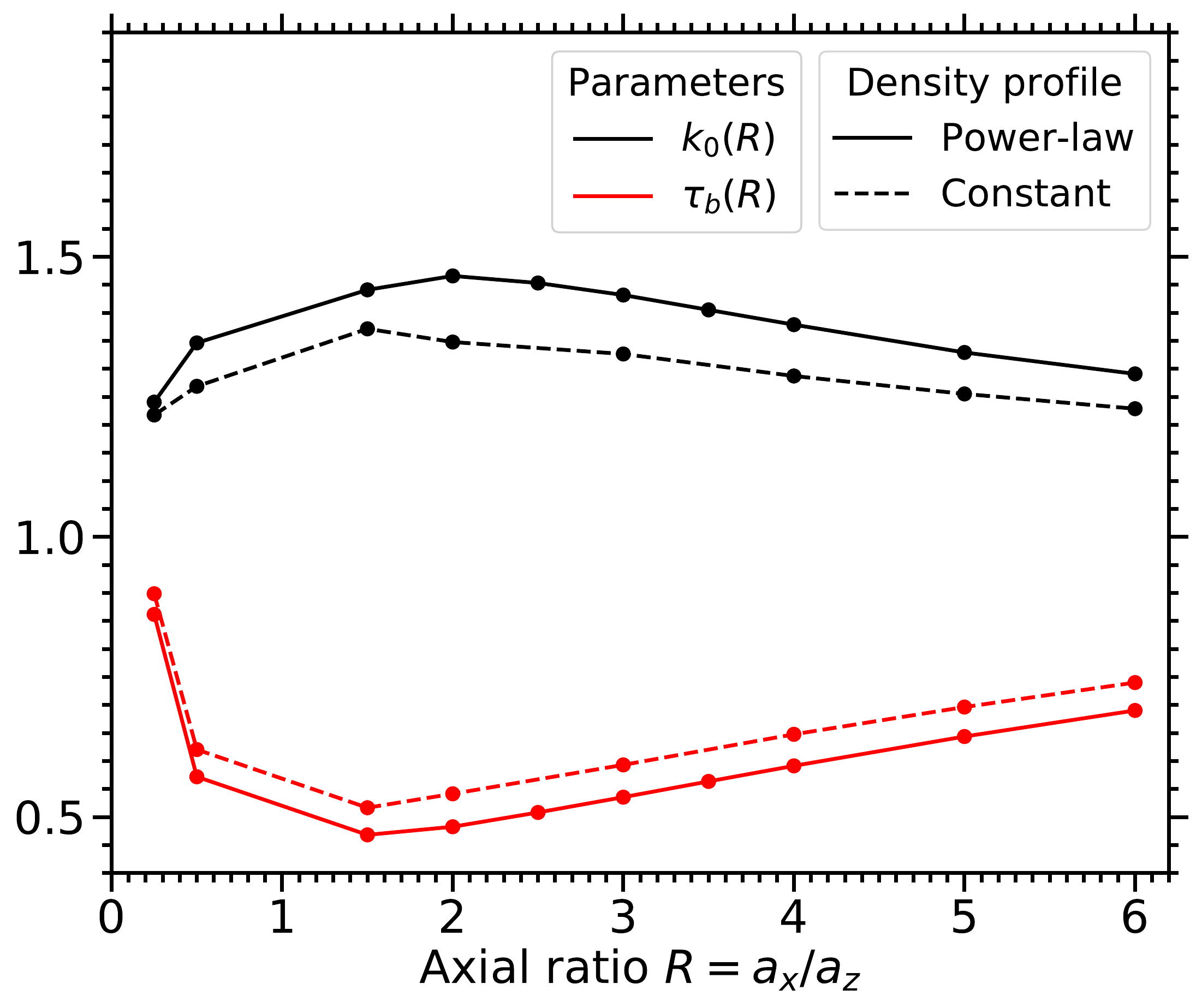}{0.33\textwidth}{(c) }
}
\caption{(a,b) The fitting parameter $k(\tau;R)$ in Equation \ref{eq:L_of_tau_mu_R_analytic} for an ellipsoid outflow with the two density profiles studied in the paper (Section \ref{subsec:methods_geometries}): (a) a broken power-law and (b) a constant. The colors show different axial ratios $R = a_x / a_z$. The time $t$ has been scaled by the diffusion time $t_d$ as $\tau = t / t_d$ (Section \ref{subsec:methods_general}). The dashed black vertical curve shows the time to peak of the light curve of the equivalent spherical ejecta ($R=1$ with the same $\tau_e$). The dashed colored curves show the fits to the analytic approximation given in Equation \ref{eq:fitting_function_k} and the bottom panels show the residuals of the fits, with (c) the parameters $k_0(R)$ (black) and $\tau_b(R)$ (red), for the broken power-law (solid) and constant (dashed) density profiles.}
\label{fig:k_vs_tau-ellipsoid}
\end{figure*}

Figure \ref{fig:k_vs_tau-ellipsoid} shows the fitting parameter $k(\tau;R)$ from Equation \ref{eq:L_of_tau_mu_R_analytic} for an ellipsoid outflow over our range $R \in [0.25, 6]$. The curves have a similar behavior for different $R$. They rise to peak values $> 1$ at times $\tau \lesssim 1$ and then decay to zero roughly exponentially. The ejecta with higher $R$ take longer to converge. Figure \ref{fig:k_vs_tau-ellipsoid} also shows the fitting parameters $k_0(R)$ and $\tau_b(R)$ from the analytic approximation to $k(\tau;R)$ given in Equation \ref{eq:fitting_function_k}. We find that the constant values $k_0(R) \simeq 1.2$ and $\tau_b(R) \simeq 0.7$ yield semi-analytic light curves with errors $\epsilon < 0.2$ for $0.25 \leq R \leq 2$ rising to $\epsilon < 0.6$ for $R = 6$.

The light curves $L(\tau,\mu;R) > 0$, and thus $k(\tau;R)$ must lie in the range $0 \leq k(\tau;R) < k_\mathrm{upper}(\tau;R)$, where $k_\mathrm{upper}(\tau;R) = ( 1 - \min_\mu A_\mathrm{proj}(\mu;R) / A_\mathrm{proj}(\mu_\mathrm{ref};R) )^{-1}$. The lower bound ensures that the semi-analytic approximation for $L(\tau,\mu;R)$ in Equation \ref{eq:L_of_tau_mu_R_analytic} satisfies $L(\tau,\mu_1;R) > L(\tau,\mu_2;R)$ when $A_\mathrm{proj}(\mu_1;R) > A_\mathrm{proj}(\mu_2;R)$, and the upper bound ensures that it satisfies $L(\tau,\mu;R) > 0$ when $A_\mathrm{proj}(\mu;R) / A_\mathrm{proj}(\mu_\mathrm{ref};R) < 1$.

An analytic approximation for $k(\tau;R)$ should remain within the above bounds as well. However, the expression given in Equation \ref{eq:fitting_function_k} increases monotonically with decreasing $\tau$, and can potentially surpass $k_\mathrm{upper}(\tau;R)$ at early times if the fitting parameters $k_0(R)$ and $\tau_b(R)$ are chosen carelessly, and thereby produce negative values for $L(\tau,\mu;R)$. This additional constraint can potentially restrict the range of $R$ over which the analytic approximation is accurate. The values for $k_0(R)$ and $\tau_b(R)$ given above preserve a positive value for $L(\tau,\mu;R)$.

\begin{figure*}
\gridline{
\fig{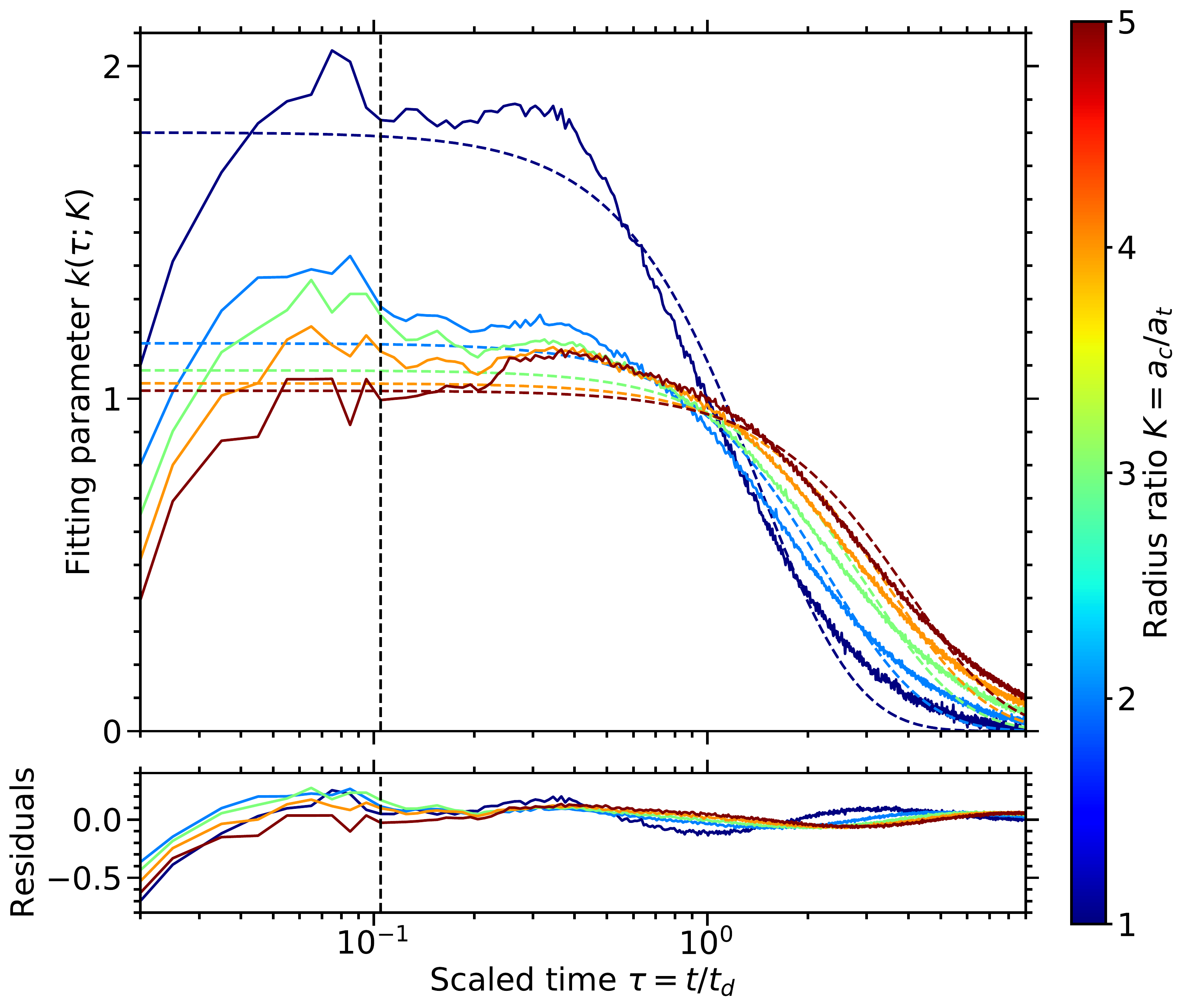}{0.32\textwidth}{(a) }
\fig{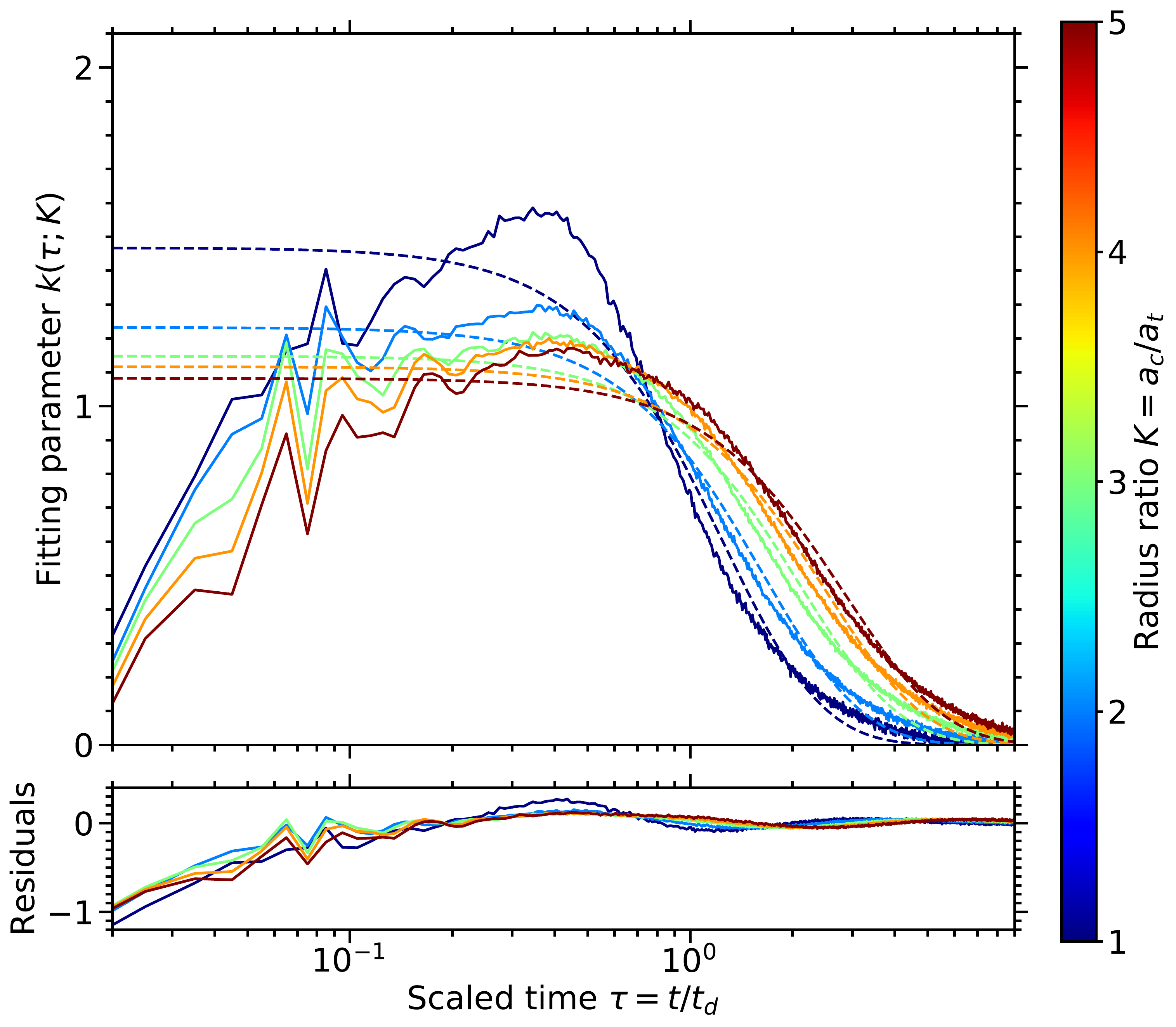}{0.32\textwidth}{(b) }
\fig{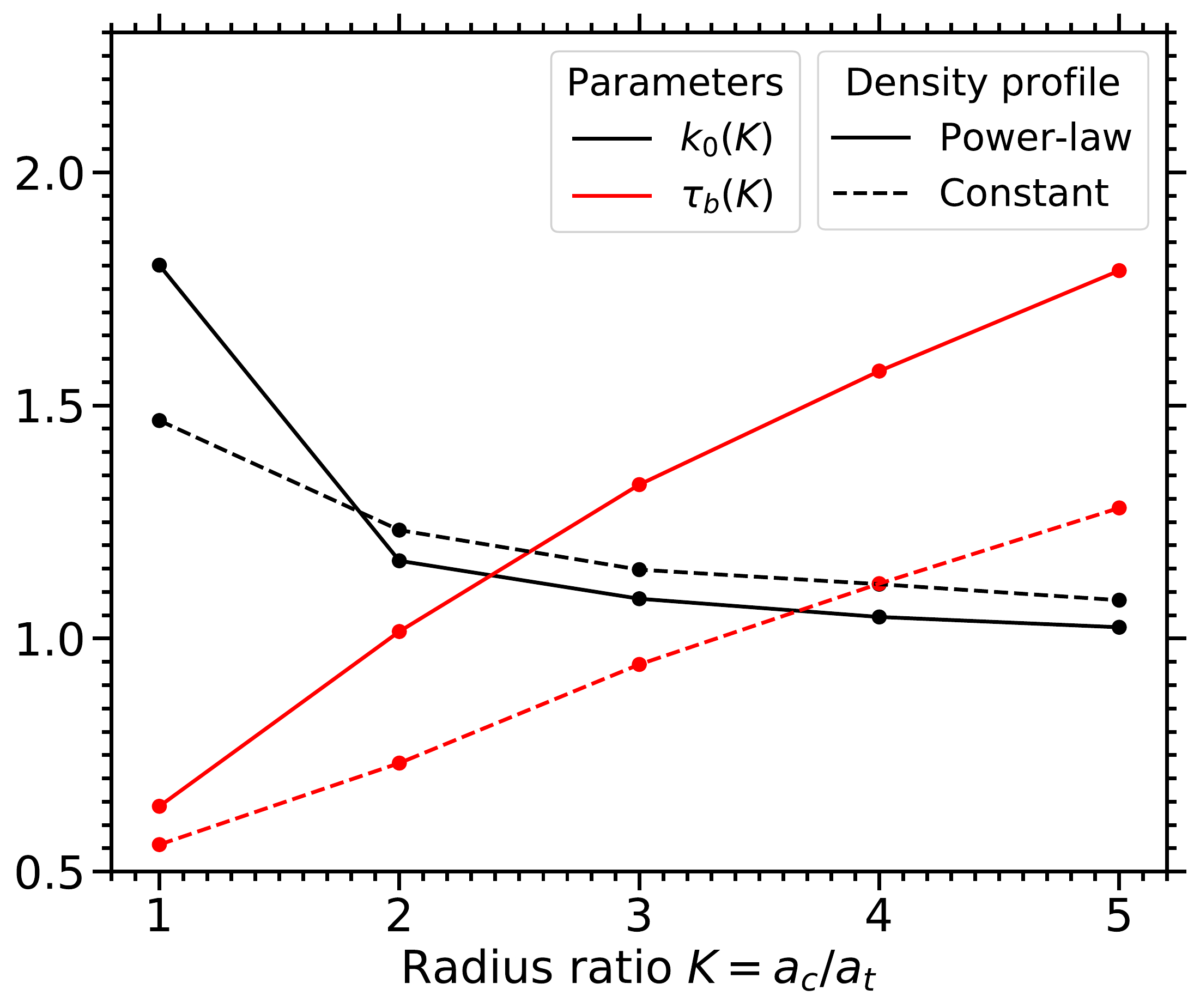}{0.33\textwidth}{(c) }
}
\caption{
(a,b) The fitting parameter $k(\tau;K)$ in Equation \ref{eq:L_of_tau_mu_R_analytic} for a torus outflow with the two density profiles studied in the paper (Section \ref{subsec:methods_geometries}): (a) a broken power-law and (b) a constant. The colors show different radius ratios $K = a_c / a_t$. The time $t$ has been scaled by the diffusion time $t_d$ as $\tau = t / t_d$ (Section \ref{subsec:methods_general}). The dashed black vertical curve shows the time to peak of the light curve of the equivalent spherical ejecta (with the same $M$, $E_k$, and $\tau_e$). The dashed colored curves show the fits to the analytic approximation given in Equation \ref{eq:fitting_function_k} and the bottom panels show the residuals of the fits, with (c) the parameters $k_0(K)$ (black) and $\tau_b(K)$ (red), for the broken power-law (solid) and constant (dashed) density profiles.
}
\label{fig:k_vs_tau-torus}
\end{figure*}

The torus outflow has analogous results, shown in Figure \ref{fig:k_vs_tau-torus}.

\subsection{Comparison of $L_\mathrm{av}$ and $L_s$}
\label{subsec:additional_features_g}

\begin{figure*}
\gridline{
\fig{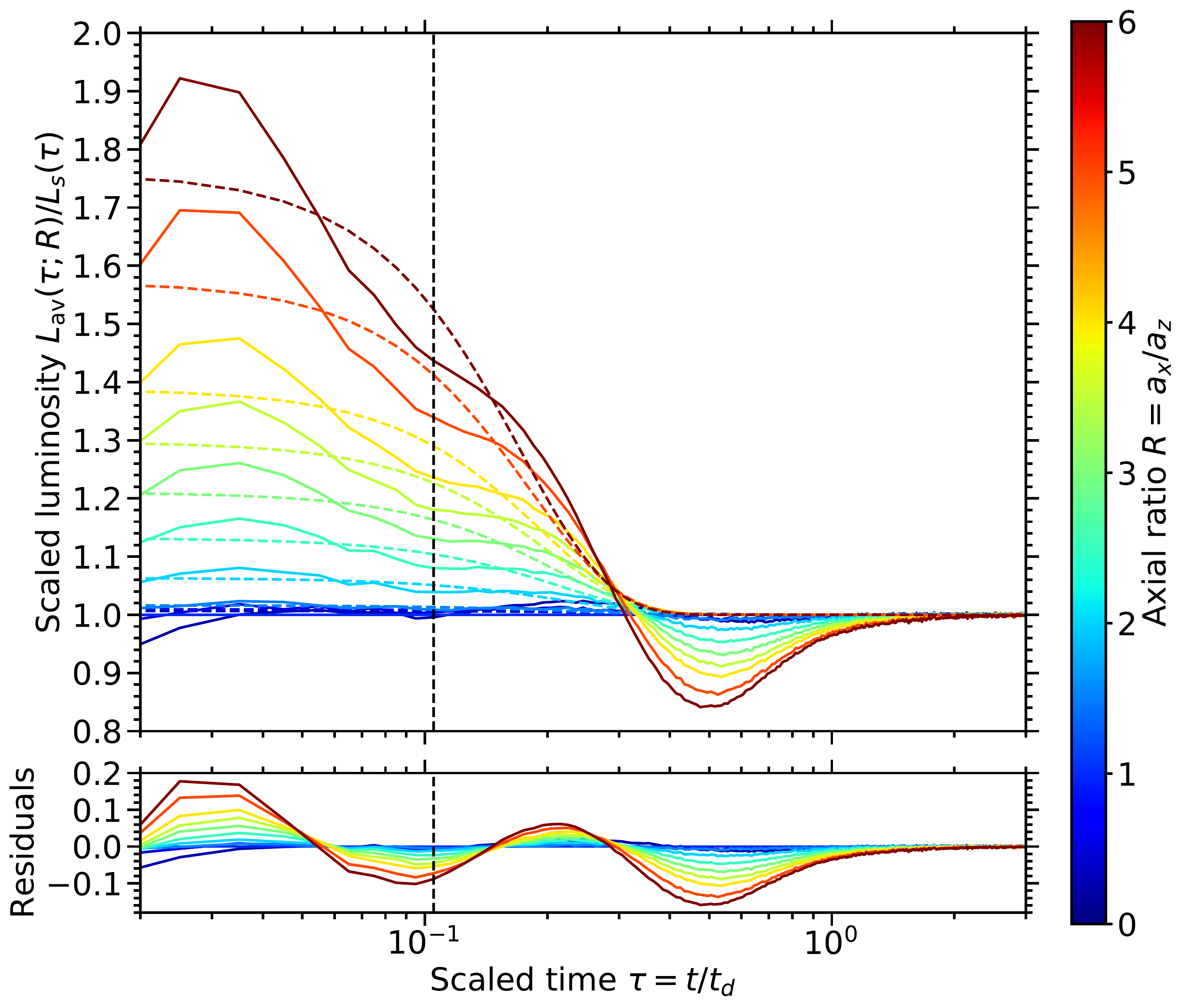}{0.32\textwidth}{(a) }
\fig{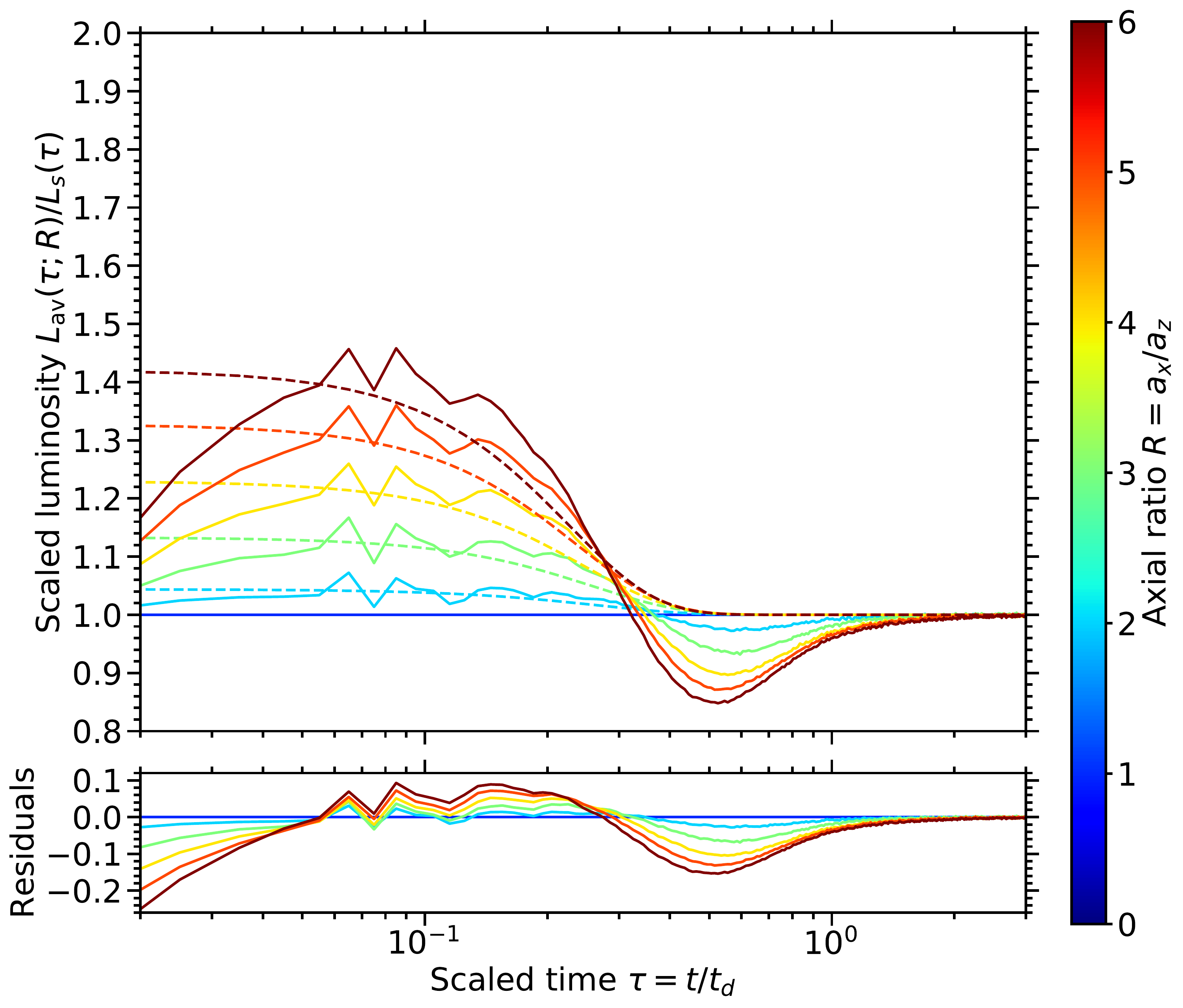}{0.32\textwidth}{(b) }
\fig{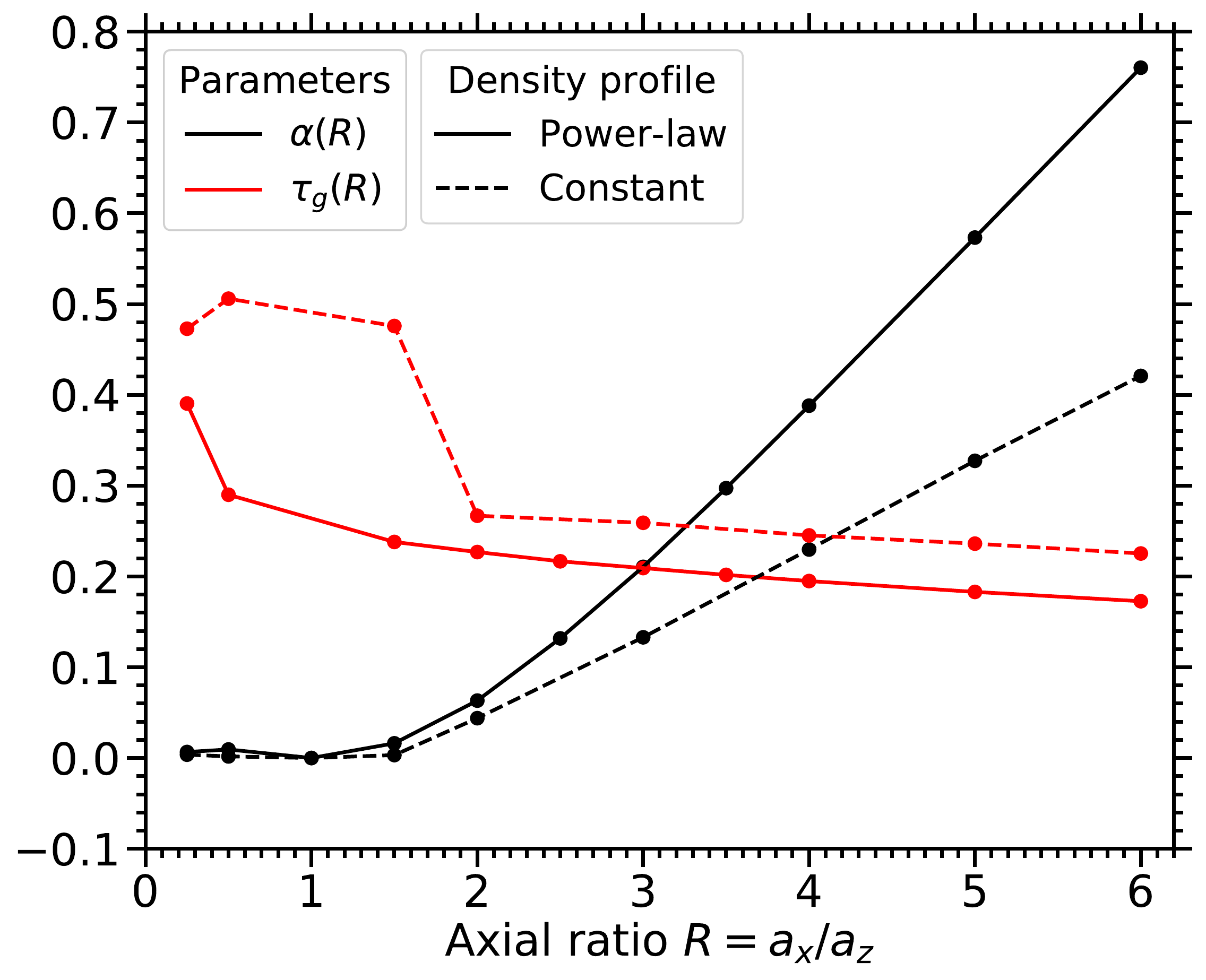}{0.33\textwidth}{(c) }
}
\caption{(a,b) The angle-averaged bolometric light curves for an ellipsoid outflow with the two density profiles studied in the paper (Section \ref{subsec:methods_geometries}): (a) a broken power-law and (b) a constant. The colors show different axial ratios $R = a_x / a_z$. The time $t$ has been scaled by the diffusion time $t_d$ as $\tau = t / t_d$ (Section \ref{subsec:methods_general}), and the angle-averaged luminosity $L_\mathrm{av}(\tau;R)$ has been scaled by the luminosity $L_s(\tau)$ of the equivalent spherical ejecta ($R=1$ with the same $\tau_e$). The dashed black vertical curve shows the time to peak of the light curve of the equivalent spherical ejecta. The dashed colored curves show the fits to the function in Equation \ref{eq:L_of_tau_R_analytic} and the bottom panel shows the residuals of the fits, with (c) the parameters $\alpha(R)$ (black) and $\tau_g(R)$ (red), for the broken power-law (solid) and constant (dashed) density profiles.}
\label{fig:g_of_tau_vs_tau-ellipsoid}
\end{figure*}

Figure \ref{fig:g_of_tau_vs_tau-ellipsoid} shows the curves $L_\mathrm{av}(\tau;R)/L_s(\tau)$ for an ellipsoid outflow over our range $R \in [0.25, 6]$, which we can examine to quantify the deviation of $L_\mathrm{av}(\tau;R)$ from $L_s(\tau)$. There is a time $\tau_\mathrm{cr}(R)$ at which $L_\mathrm{av}(\tau;R) / L_s(\tau) = 1$, i.e. where $L_\mathrm{av}(\tau;R)$ and $L_s(\tau)$ cross, and it falls in the range $0.3 \leq \tau_\mathrm{cr}(R) \leq 0.37$. For $\tau > \tau_\mathrm{cr}(R)$, $L_\mathrm{av}(\tau;R)$ and $L_s(\tau)$ differ by an error $\epsilon < 0.2$, and we can roughly take $L(\tau;R) / L_s(\tau) \simeq 1$. For $\tau < \tau_\mathrm{cr}(R)$, $L_\mathrm{av}(\tau;R)$ is larger than $L_s(\tau)$, and the deviation is greater for ejecta with higher $R$. This is because photons can escape more easily in the $z$-direction for the ellipsoidal ejecta, leading to a lower effective diffusion time. In this region, we can roughly take $L_\mathrm{av}(\tau;R) / L_s(\tau) \simeq 1$ for $R \lesssim 3$ since we only incur an error $\epsilon < 0.2$, but should use a more accurate expression for $R \gtrsim 3$. It is important to be accurate in this region since the light curves experience their peak here.

\begin{figure*}
\gridline{
\fig{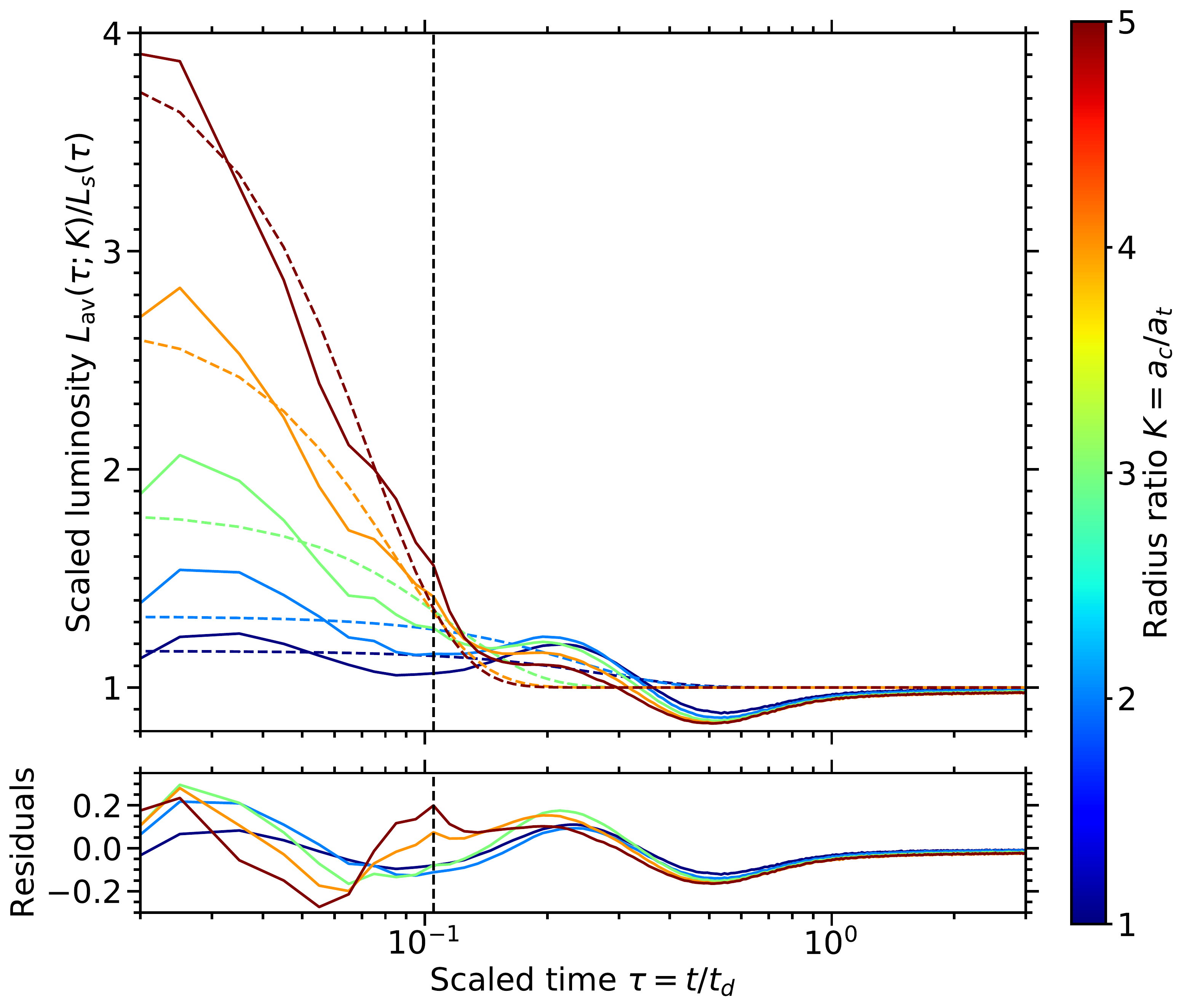}{0.32\textwidth}{(a) }
\fig{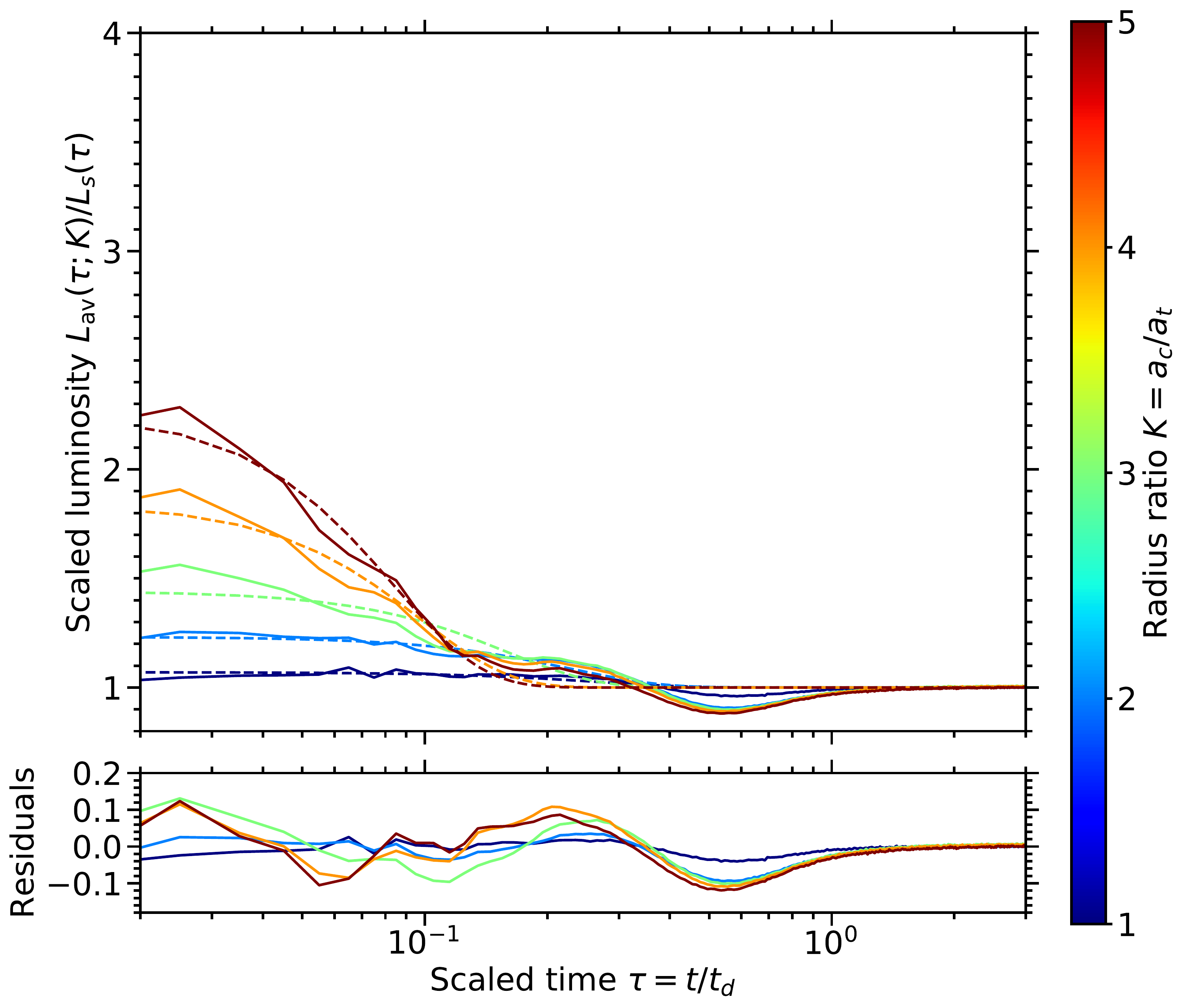}{0.32\textwidth}{(b) }
\fig{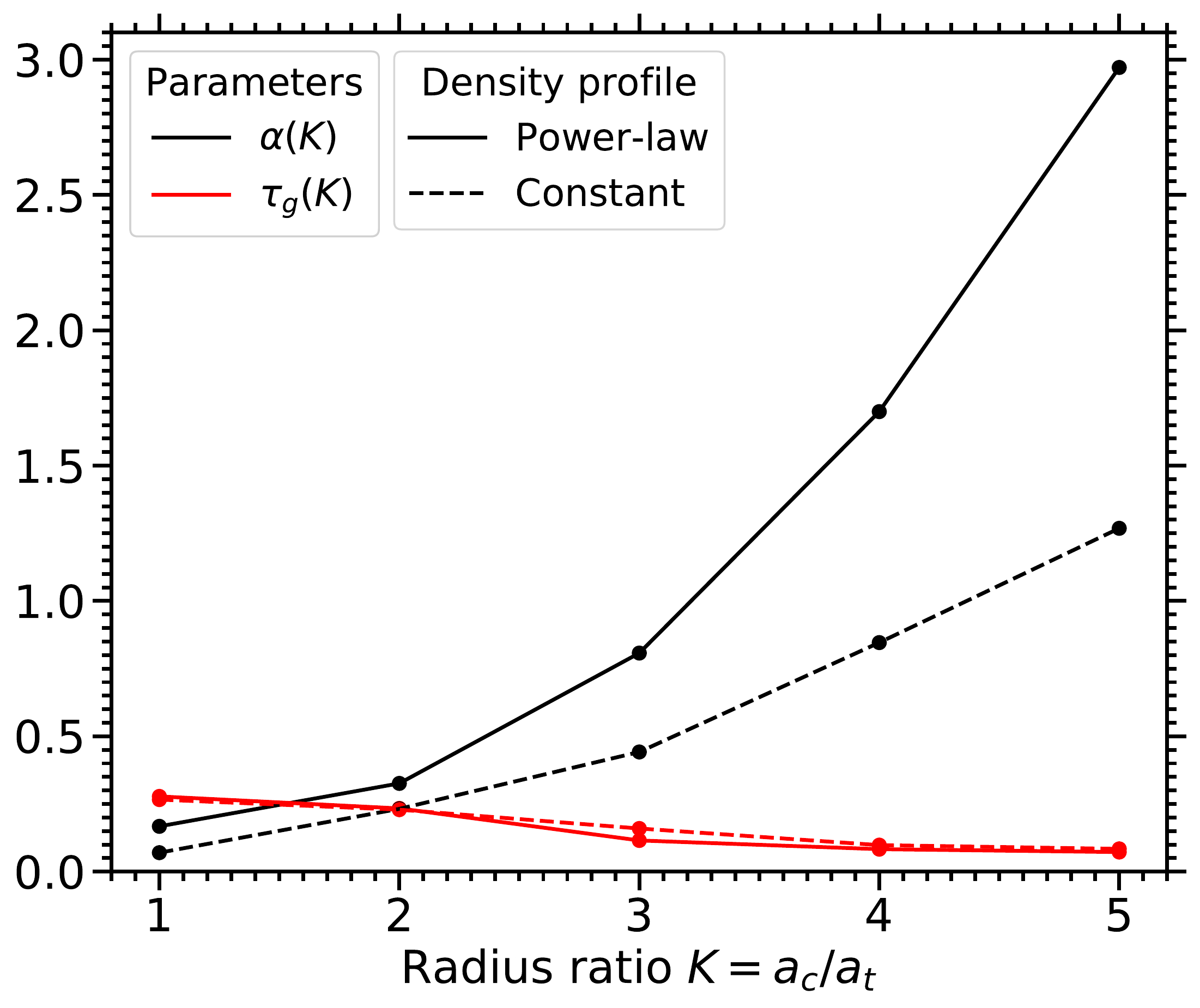}{0.33\textwidth}{(c) }
}
\caption{
(a,b) The angle-averaged bolometric light curves for a torus outflow with the two density profiles studied in the paper (Section \ref{subsec:methods_geometries}): (a) a broken power-law and (b) a constant. The colors show different radius ratios $K = a_c / a_t$. The time $t$ has been scaled by the diffusion time $t_d$ as $\tau = t / t_d$ (Section \ref{subsec:methods_general}), and the angle-averaged luminosity $L_\mathrm{av}(\tau;K)$ has been scaled by the luminosity $L_s(\tau)$ of the equivalent spherical ejecta (with the same $M$, $E_k$, and $\tau_e$). The dashed black vertical curve shows the time to peak of the light curve of the equivalent spherical ejecta. The dashed colored curves show the fits to the function in Equation \ref{eq:L_of_tau_R_analytic} and the bottom panel shows the residuals of the fits, with (c) the parameters $\alpha(K)$ (black) and $\tau_g(K)$ (red), for the broken power-law (solid) and constant (dashed) density profiles.
}
\label{fig:g_of_tau_vs_tau-torus}
\end{figure*}

Motivated by this behavior, we adopt the following analytic approximation
\begin{equation}
\frac{L_\mathrm{av}(\tau;R)}{L_s(\tau)} \simeq 1 + \alpha(R) e^{-\tau^2/\tau_g^2(R)}
\label{eq:L_of_tau_R_analytic}
\end{equation}
Here, $\alpha(R)$ is the amplitude and $\tau_g(R)$ is the decay time, which falls in the range $\tau_g(R) \lesssim \tau_\mathrm{cr}(R)$. The function thus provides a fit in the region $\tau \lesssim \tau_\mathrm{cr}(R)$, and is $L_\mathrm{av}(\tau;R)/L_s(\tau) \simeq 1$ in the region $\tau \gtrsim \tau_\mathrm{cr}(R)$. Figure \ref{fig:g_of_tau_vs_tau-ellipsoid} shows the fitting parameters $\alpha(R)$ and $\tau_g(R)$ obtained from fitting the curves $L_\mathrm{av}(\tau;R)/L_s(\tau)$ with $g(\tau;R)$. For $R \lesssim 3$, we can approximate $L_\mathrm{av}(\tau;R)/L_s(\tau) \simeq 1$, incurring an error of $\epsilon \lesssim 0.2$ for all $\tau$. For $R \gtrsim 3$, one should use the full expression for $L_\mathrm{av}(\tau;R)/L_s(\tau)$ if an error $\lesssim 0.2$ is needed.

We can thus obtain a more accurate projection factor from $L_s(\tau)$ to $L(\tau,\mu;R)$ by combining Equation \ref{eq:L_of_tau_mu_R_analytic} (with the replacement $L(\tau,\mu_\mathrm{ref};R) \rightarrow L_\mathrm{av}(\tau;R)$), Equation \ref{eq:L_of_tau_R_analytic}, and Equation \ref{eq:fitting_function_k}. This parameterization is more involved. The numerical models are also available if accurate light curves are needed.

The torus outflow has analogous results, shown in Figure \ref{fig:g_of_tau_vs_tau-torus}.

\bibliographystyle{aasjournal}
\bibliography{references} % if your bibtex file is called references.bib

%% This command is needed to show the entire author+affilation list when
%% the collaboration and author truncation commands are used.  It has to
%% go at the end of the manuscript.
%\allauthors

%% Include this line if you are using the \added, \replaced, \deleted
%% commands to see a summary list of all changes at the end of the article.
%\listofchanges

\end{document}